\documentclass[pdflatex,sn-basic,Numbered]{sn-jnl}
\usepackage[T1]{fontenc}
\usepackage[utf8]{inputenc}
\usepackage{amsmath,amssymb,mathtools}
\usepackage{xcolor}
\usepackage{graphicx,longtable,booktabs,array,calc,multirow}
\usepackage{adjustbox}
\usepackage{microtype}
\usepackage{xurl}
\usepackage{fancyvrb}
\usepackage{caption}
\usepackage{placeins}
\usepackage{needspace}
\usepackage{etoolbox}
\usepackage{bookmark}
\hypersetup{hidelinks,pdftitle={Direct cognition problem: a semantic skeleton of mathematics and its growth},pdfauthor={Vasilii A. Gromov; Nikolai N. Evdokimov}}
\newcommand{\tightlist}{\setlength{\itemsep}{0pt}\setlength{\parskip}{0pt}}

\newsavebox{\displaybox}
\newcommand{\fitdisplay}[1]{\sbox{\displaybox}{$\displaystyle #1$}\ifdim\wd\displaybox>\linewidth\resizebox{\linewidth}{!}{\usebox{\displaybox}}\else\usebox{\displaybox}\fi}
\makeatletter
\newsavebox\pandoc@box
\newcommand*\pandocbounded[1]{\sbox\pandoc@box{#1}\ifdim\wd\pandoc@box>\linewidth\resizebox{\linewidth}{!}{\usebox\pandoc@box}\else\usebox\pandoc@box\fi}
\patchcmd\longtable{\par}{\if@noskipsec\mbox{}\fi\par}{}{}
\makeatother
\begin{document}
\title[Direct cognition problem: a semantic skeleton of mathematics and its growth]{Direct cognition problem: a semantic skeleton of mathematics and its growth}
\subtitle{Epochs limit active premises while certificates preserve proofs}
\author[1]{\fnm{Vasilii A.} \sur{Gromov}}\email{stroller@rambler.ru}
\author*[1]{\fnm{Nikolai N.} \sur{Evdokimov}}\email{nevdokimov@hse.ru}
\affil[1]{\orgdiv{Laboratory for Semantic Analysis}, \orgname{National Research University Higher School of Economics}, \orgaddress{\street{11 Pokrovsky Bulvar}, \city{Moscow}, \postcode{109028}, \country{Russia}}}
\abstract{We construct a finite semantic skeleton from three axioms of classical propositional logic without specifying a target theorem. During growth, the constructor records which proofs share an intermediate lemma. When choosing the final anchors, it also counts inferences that use a premise the lemma can supply by substitution. The primary skeleton contains ten anchors whose certificates expand to the axioms through 27 shared inference steps. It shortens all 30 previously unseen proof structures in an external catalogue of typed combinators, saving 90 of 412 steps. One anchor recurs in all five runs of the constructor. In a post hoc test, this anchor alone shortens 26 of the 30 proofs. It also shortens 331 of 475 previously unseen proofs of familiar theorems in a second corpus.}
\maketitle
\noindent\small ORCID (Nikolai N. Evdokimov): \url{https://orcid.org/0009-0003-3756-8281}\par\normalsize

\section*{1. Introduction}\label{introduction}
\addcontentsline{toc}{section}{1. Introduction}

Mathematicians use some proved statements as stepping stones in many arguments. A formal proof library preserves these statements alongside thousands of others. Which of them should a small reusable library retain? We use \emph{semantic skeleton} for a system of anchor statements and the proof dependencies that let a checker recover them. The term refers to the roles that statements play in deductions under a specified calculus. Classical propositional logic admits infinitely many theorem schemata. A run of our constructor yields one finite fragment of this larger structure.

We distinguish three cognition problems. In the \textbf{direct problem}, we grow a proof skeleton without a target theorem. We seek proved statements that can support further deductions under limited resources.\footnote{Bachmair and Ganzinger \citep{BachmairGanzinger2001}, in \emph{Resolution Theorem Proving}, describe how saturation combines inference rules with criteria for discarding redundant clauses. Wernhard \citep{Wernhard2026}, in \emph{Generating Theorems by Generating Proof Structures}, generates theorems without a target. Section 2 explains how these methods relate to ours.} In the \textbf{inverse problem}, we start with a particular statement and use it to decide where to extend the skeleton. Intermediate lemmas can connect a distant target to existing anchors over several steps. A proof of the statement, a proof of its negation, or a separate independence argument can resolve its status relative to the chosen axioms. If the search exhausts its budget, its answer remains unknown. In the \textbf{control problem}, we choose how the graph grows to obtain properties such as modularity, concentrated reuse, or a prescribed degree distribution. Section~9.1 returns to the inverse problem and outlines a test on long proofs in Lean.

We give a skeleton with ten anchors. Starting from three axiom schemata, the constructor grows proofs forward and records where they share an intermediate conclusion. When it selects the final anchors, it also checks whether a schema can supply a premise in those proofs by substitution. A reader can cite the schema in another proof and expand the citation to the original axioms. Appendix I gives all ten schemata and their shared certificate.

We test the skeleton against one constructed from Wernhard's SGCD archive under the same limits and the same rule for choosing anchors. We report how many external proofs each skeleton shortens, how many distinct inference steps it saves, and what its certificates cost once. Separate experiments examine what the constructor can derive if growth continues and how the choice of proof edges affects its graph.

Section 2 reviews related work. Section 3 defines the skeleton and the criteria for testing it. Section 4 bounds the size of the explicit theorem space. Section 5 describes the generator, and Section 6 sets out the test protocols. Section 7 reports the results, Section 8 discusses their scope, and Section 9 draws conclusions and outlines an application in Lean. Section 10 explains how to trace the results. Appendices A--I supply proofs, protocols, and the primary skeleton.

\section*{2. Related papers}\label{related-papers}
\addcontentsline{toc}{section}{2. Related papers}

Work on automated theory exploration supplies methods for discovering theorems and useful lemmas. Work on inference control shows how provers choose which statements to process. We draw on both strands to separate the cost of storing proofs, the cost of keeping premises active, and the work that those premises support during subsequent search.

\subsection*{2.1 Saturation and search-resource limits}\label{saturation-and-search-resource-limits}
\addcontentsline{toc}{subsection}{2.1 Saturation and search-resource limits}

Bachmair and Ganzinger (\citeyear{BachmairGanzinger2001}), in their handbook chapter \emph{Resolution Theorem Proving}, explain how saturation combines inference rules with redundancy criteria. Schulz (\citeyear{Schulz2002}), in \emph{E --- A Brainiac Theorem Prover}, and Kov\'{a}cs and Voronkov (\citeyear{KovacsVoronkov2013}), in \emph{First-Order Theorem Proving and Vampire}, describe how clause indices, scores, and simplification direct that search. Target-free forward inference therefore supplies background rather than our novelty claim. We retain a long-lived archive of checked consequences and assess the skeleton through new growth under bounded activity.

Riazanov and Voronkov (\citeyear{RiazanovVoronkov2003LRS}), in \emph{Limited Resource Strategy in Resolution Theorem Proving}, adapt permissible clause weights to remaining time and search statistics. Their strategy directly precedes resource-based pruning. Logical redundancy and low expected computational return support different guarantees: a suitable redundancy criterion can preserve completeness, whereas a heuristic return estimate need not. Our age-based pruning protects earlier certificates but does not ensure fair future enumeration.

Hoder and Voronkov (\citeyear{HoderVoronkov2011}), in \emph{Sine Qua Non for Large Theory Reasoning}, reduce the initial theory for a particular problem through SInE. Blanchette and colleagues (\citeyear{BlanchetteEtAl2016}), in \emph{A Learning-Based Fact Selector for Isabelle/HOL}, learn fact relevance from earlier proof dependencies. Jakub\r{u}v and Urban (\citeyear{JakubuvUrban2017}), in \emph{ENIGMA: Efficient Learning-Based Inference Guiding Machine}, rank clauses during search according to how they contribute to successful proofs. These studies distinguish how a prover initially chooses premises from how it controls search. We additionally distinguish storing a fragment from admitting its conclusion to a bounded active slot, without labelling success against a final target.

\subsection*{2.2 Theory exploration and the usefulness of new statements}\label{theory-exploration-and-the-usefulness-of-new-statements}
\addcontentsline{toc}{subsection}{2.2 Theory exploration and the usefulness of new statements}

Colton (\citeyear{Colton2002}), in his monograph \emph{Automated Theory Formation in Pure Mathematics}, develops automated concept and conjecture discovery. Smallbone and colleagues (\citeyear{SmallboneEtAl2017}), in \emph{Quick Specifications for the Busy Programmer}, present QuickSpec, which enumerates terms, tests conjectural equalities, and reduces redundancy through their dependencies. Claessen and colleagues (\citeyear{ClaessenEtAl2013}), in \emph{Automating Inductive Proofs Using Theory Exploration}, combine such exploration with inductive proof in HipSpec. These systems show how auxiliary statements can support later work; our archive instead accepts each statement together with its checked derivation.

Johansson and colleagues (\citeyear{JohanssonEtAl2014}), in \emph{Hipster: Integrating Theory Exploration in a Proof Assistant}, validate conjectures through Isabelle/HOL and assess interestingness against ordinary and stronger proof mechanisms. Singher and Itzhaky (\citeyear{SingherItzhaky2021}), in \emph{Theory Exploration Powered by Deductive Synthesis}, combine symbolic enumeration, rewriting, congruence closure, and inductive checking in TheSy. Their bounded solver also examines mutual derivability. These studies motivate separate measures of novelty, proof status, and usefulness; our archive checks every accepted derivation, whereas our experiments assess continued growth rather than mutual derivability of complete theories.

McCasland, Bundy, and Smith (\citeyear{McCaslandBundySmith2017}), in MATHsAiD: Automated mathematical theory exploration, describe theorem discovery from supplied axioms and definitions and distinguish it from proving a supplied goal. This distinction also frames our direct problem. We distinguish archived certificates, reusable fragments, and active premises in our experiments, and measure how much continued generation costs.

\subsection*{2.3 Generating proof structures: the closest predecessors}\label{generating-proof-structures-the-closest-predecessors}
\addcontentsline{toc}{subsection}{2.3 Generating proof structures: the closest predecessors}

Wernhard and Bibel (\citeyear{WernhardBibel2024}), in \emph{Investigations into Proof Structures}, distinguish proof-tree size from minimal directed acyclic proof-graph size and study level-wise enumeration, theorem-size bounds, and bounded caches. Their analysis directly informs our generator. We extend the experimental focus to persistent certified growth and separately control archived proofs, reusable fragments, and active premises.

Wernhard (\citeyear{Wernhard2026}) generates theorems by enumerating proof structures, synthesises lemmas, and compares the output with human Metamath theorems. We use his SGCD generator as a direct comparator. Both generators start from the same three axiom schemata. We extract anchors from each archive and test the two resulting skeletons on the same external proofs.

Rawson and colleagues (\citeyear{RawsonEtAl2023}), in \emph{Lemmas: Generation, Selection, Application}, separate those three stages and compare structural usefulness scores with how proving work changes after they add lemmas. Their tests motivate our distinction between shorter proof code and faster subsequent search. They assess lemmas through proof tasks; our principal trajectories instead continue without an external final target.

Wernhard and Zombori (\citeyear{WernhardZombori2025}), in their preprint \emph{Mathematical Knowledge Bases as Grammar-Compressed Proof Terms: Exploring Metamath Proof Structures}, encode libraries through compressed proof terms, including parametrised fragments. Our code-length criterion scores particular closed first proofs, whereas a separate compiler derives general closed schemata from open modus-ponens programmes. The archive already shares subproofs. Consequently, shorter expanded forests measure the declared proof-term code rather than additional compression of the archive graph or universal value of lemmas for human mathematics.

\subsection*{2.4 Acquired knowledge: from recurrence to computational return}\label{acquired-knowledge-from-recurrence-to-computational-return}
\addcontentsline{toc}{subsection}{2.4 Acquired knowledge: from recurrence to computational return}

Minton (\citeyear{Minton1990Utility}), in \emph{Quantitative Results Concerning the Utility of Explanation-Based Learning}, measures how applicability checks and maintenance can outweigh gains from acquired knowledge. In our setting, premises compete for a bounded number of active slots. Frequent fragment occurrences establish reuse, but only a costed continuation test can assess whether access to those conclusions yields more new results. We therefore report code-length savings separately from computational return.

Kaliszyk and Urban (\citeyear{KaliszykUrban2015}), in \emph{Learning-Assisted Theorem Proving with Millions of Lemmas}, select intermediate lemmas through acquisition cost, reuse, centrality, and greedy changes to the dependency graph. They also distinguish protocols by access to future proofs. Our selector recalculates overlaps using only the preceding forest, because earlier choices change the value of another fragment. Development diagnostics measure human-statement coverage after a run; a separate locked comparison evaluates a previously unscored target pool.

\subsection*{2.5 Trainable generators and growing libraries}\label{trainable-generators-and-growing-libraries}
\addcontentsline{toc}{subsection}{2.5 Trainable generators and growing libraries}

Wang and colleagues (\citeyear{WangEtAl2023Lego}), in \emph{LEGO-Prover: Neural Theorem Proving with Growing Libraries}, build and reuse verified auxiliary skills while solving formal tasks. Yang and colleagues (\citeyear{YangEtAl2023}), in \emph{LeanDojo: Theorem Proving with Retrieval-Augmented Language Models}, combine premise retrieval with search and track fact availability. These systems inform future uses of a skeleton for target-driven reasoning. The propositional and modal growth panels start from axioms, whereas the native extensions inherit a certified source library. Neither calls a language model for formulae or fragments.

Poesia and colleagues (\citeyear{PoesiaEtAl2024}), in \emph{Learning Formal Mathematics from Intrinsic Motivation}, jointly learn conjecturing and proving in MINIMO from an initial axiomatisation. Ota and colleagues (\citeyear{OtaEtAl2026}), in their preprint \emph{Self-Supervised Theorem Discovery in a Formal Axiomatic System}, turn discovered theorems into prover actions and select them through generalisation and difficulty of rediscovery. Both systems create their own search targets. Our generator instead continues forward without a trainable target-setting loop.

Lin and colleagues (\citeyear{LinEtAl2024}), in \emph{ATG: Benchmarking Automated Theorem Generation for Generative Language Models}, assess generated theorems partly through their usefulness for subsequent proving. We test neutral direct continuation and additional coverage of human statements. This choice measures further target-free growth; independent goal-directed tasks would test the complementary inverse problem.

\subsection*{2.6 Prover cooperation and graphs of mathematical libraries}\label{prover-cooperation-and-graphs-of-mathematical-libraries}
\addcontentsline{toc}{subsection}{2.6 Prover cooperation and graphs of mathematical libraries}

Tomaszuk and colleagues (\citeyear{TomaszukEtAl2023}), in \emph{MMLKG: Knowledge Graph for Mathematical Definitions, Statements and Proofs}, represent the existing Mizar library as a knowledge graph. Li and colleagues (\citeyear{LiEtAl2026}), in their preprint \emph{The Network Structure of Mathlib}, distinguish dependency levels and relate observed centrality to formalisation infrastructure. These studies motivate a graph-level account of mathematical libraries. Our generator additionally requires joint premises and expandable certificates; high degree, short paths, or a compact projection cannot replace those proof conditions.

\subsection*{2.7 Position of this work}\label{position-of-this-work}
\addcontentsline{toc}{subsection}{2.7 Position of this work}

The object we compare is a set of checked anchor schemata together with their shared proof support. Any anchor can act as a premise under an admissible substitution. Its archived dependencies explain how to recover the proof if another system lacks that anchor. The external test asks how much work a reference to such a schema saves at an intermediate step of a proof.

\section*{3 Formal model and outcome criteria}\label{formal-model-and-outcome-criteria}
\addcontentsline{toc}{section}{3 Formal model and outcome criteria}

\subsection*{3.1 Calculus and proof state}\label{calculus-and-proof-state}
\addcontentsline{toc}{subsection}{3.1 Calculus and proof state}

A proof certificate records elementary inferences and substitutions that a checker can replay from the initial base. We call a statement proven once its certificate passes those checks, and active when the generator may use it as a premise. A schema contains variables that stand for admissible expressions; an instance replaces those variables consistently. A configuration fixes the language, proof rules, and contextual requirements:

\[\fitdisplay{\mathcal{C} = (\Sigma,\Gamma,\Re,\nu,\Xi),}\]

where \(\Sigma\) specifies the language, \(\Gamma\) denotes the initial axioms or schemata, \(\Re\) contains the rules, \(\nu\) maps each statement to its declared canonical form, and \(\Xi\) describes contextual restrictions. In systems with quantifiers and hypotheses, each vertex includes the necessary context, variable types, and admissible-substitution conditions alongside its formula. Formula identity alone therefore need not identify the full proof state.

A typed hypergraph\footnote{A typed directed hypergraph records statements as vertices, together with their variable types, hypotheses, and side conditions. A directed hyperedge joins an ordered group of premises to one conclusion and records the rule and substitution. Its premise order identifies rule roles; the inference requires those premises jointly.

  For modus ponens (MP), the minor premise \(p\) and major premise \(p \rightarrow q\) jointly yield \(q\); neither premise alone supplies that inference.} \(H = (V,E)\) provides the primary representation. We record a rule application as

\[\fitdisplay{e = (p_{1},\ldots,p_{k};\rho;\sigma;\chi;q),}\]

where \(p_{1},\ldots,p_{k}\) list premises in the order that the rule requires, \(\rho\) denotes the rule, \(\sigma\) denotes the substitution, \(\chi\) contains side conditions, and \(q\) denotes the conclusion. Each hyperedge requires all its premises jointly, whereas alternative hyperedges represent different derivations of the same conclusion. The representation thus preserves the distinction between jointly necessary premises and alternative proofs.

To canonicalise a formula, we remove only the notational differences that the configuration specifies. In the implicational fragment, we rename variables in the order in which they first appear. A canonical identifier labels the resulting syntax. Logical equivalence alone does not merge two vertices.

\subsection*{3.2 Requirements for a certified skeleton}\label{skeleton-contract}
\addcontentsline{toc}{subsection}{3.2 Requirements for a certified skeleton}

\textbf{Definition 1.} Fix a calculus and its permitted initial base. We call a proved statement or schema an \emph{anchor} when we select it as a reusable premise and retain its certificate. A finite certified proof skeleton consists of a triple

\[\fitdisplay{\mathcal{S} = (S,\Omega,\mathsf{E}\mathsf{x}\mathsf{p}\mathsf{a}\mathsf{n}\mathsf{d}),}\]

where \(S\) lists finitely many anchors, \(\Omega\) contains their certificates and dependencies, and \(\mathsf{Expand}\) replaces an anchor reference by its proof from the initial base. Dependencies follow a well-founded order. The checker verifies the rule, substitution, and side conditions at each step. Distinct anchors may share certificate steps. The experiment fixes the calculus and a selection rule, so different finite runs can yield different valid fragments.

One expandable witness suffices for each anchor. An implementation may keep several witnesses with different costs or share subproofs across certificates. If it deletes a witness, it must keep another way to recover a checked proof for every anchor that depended on it.

For example, the proof of the typed combinator \texttt{finch} takes 16 distinct inference steps from the axioms. Three anchors supply intermediate conclusions in that proof, including an instance of \(A\to A\). With those anchors available, nine steps remain, including the final inference. Appendix I.1 gives the three node identifiers and their certificates.

\subsection*{3.3 Reconstruction and new growth}\label{reconstruction-and-new-growth}
\addcontentsline{toc}{subsection}{3.3 Reconstruction and new growth}

The model uses two operators to recover proofs and to grow the skeleton.

\[\fitdisplay{\mathsf{E}\mathsf{x}\mathsf{p}\mathsf{a}\mathsf{n}\mathsf{d}(\mathcal{S},v) \rightarrow \pi_{v},\quad\quad\mathsf{G}\mathsf{r}\mathsf{o}\mathsf{w}(\mathcal{S},b) \rightarrow \Omega_{b}.}\]

The first operator returns a certificate for an existing anchor. The second performs forward inference under budget \(b\) without a final target formula, using the initial axioms, selected anchors, and its own new consequences. Vertices that serve only as proof support do not automatically enter the available premise set, so archival presence alone does not confer search activity.

New-growth counts exclude rediscovery of formulae that the archive already contains. Let \(\Omega_{0}\) denote the common initial archive from which the comparison policies choose anchors, and define the main direct-continuation measure as

\[\fitdisplay{N_{b}(\mathcal{S} \mid \Omega_{0}) = \left| V(\Omega_{b})\setminus V(\Omega_{0}) \right|.}\]

This measure counts syntactic novelty under the declared canonicalisation, even after earlier formulae leave the active skeleton. It does not establish mathematical significance for every new result.

The native continuation tests use typed one-way coverage with mandatory distinct-variable conditions. For source block \(j\), let \(T\) denote the external reference and let \(C_{T}(S_{j,m},0)\) denote the statements that method \(m\) covers before further search. We freeze a common residual

\[\fitdisplay{R_{j} = T\setminus \bigcup_{m}C_{T}(S_{j,m},0),\quad\quad Q_{j}(m,h) = \left| R_{j} \cap C_{T}(S_{j,m},h) \right|.}\]

The observer counts how many reference statements the committed results cover. It supplies no premise, reward, or proposed conclusion. The residual removes every method's initial coverage advantage within a block. Counts of \(Q_{j}\) test new reference capability, whereas \(N_{b}\) counts syntactically new formulae outside the common archive. Neither measure substitutes for the other.

Resource accounting distinguishes the active set and reserve from the library, proof support, and archive. We measure their sizes separately, together with peak memory for the entire process; a fixed premise count does not imply fixed total memory. The principal size bounds \(L = 25,200\) count every syntax-tree vertex, not only leaves, and therefore constrain a different quantity from the active-set limit.

\subsection*{3.4 Useful skeletons and empirical Pareto sets}\label{useful-skeletons-and-empirical-pareto-sets}
\addcontentsline{toc}{subsection}{3.4 Useful skeletons and empirical Pareto sets}

For an external proof \(p\), let \(d(p)\) count distinct derived steps in its directed acyclic graph. Let \(d_S(p)\) count the steps that remain when a matching anchor in \(S\) supplies a proper intermediate conclusion. We retain the final inference even if an anchor matches the theorem itself. A shared step survives whenever any remaining branch still needs it. For a panel \(P\), we report the number of proofs with \(d_S(p)<d(p)\) and the total saving \(\sum_{p\in P}[d(p)-d_S(p)]\). The latter counts reuse across different proofs. For net savings, we subtract once the number of distinct inference steps needed to establish the entire anchor set.

For each policy and anchor cap, we record a triple: the measured cost of constructing and using a skeleton, the bytes in its proof-backed package, and the new schemata that it generates during continuation. One setting dominates another only when it uses no more CPU or package bytes, yields at least as many new schemata, and improves at least one coordinate. The empirical Pareto set contains the nondominated settings within the declared finite comparison grid. This definition avoids assigning an arbitrary exchange rate between a byte, a CPU-second, and a new formula.

We compare settings only within one logic--size regime, one source ceiling, one continuation horizon, and a common exclusion archive. Package bytes include both anchors and certificate support; an anchor count alone omits that support. 

\section*{4 Limits of exhaustive explicit construction}\label{limits-of-exhaustive-explicit-construction}
\addcontentsline{toc}{section}{4 Limits of exhaustive explicit construction}

The combinatorial argument depends on the unit of storage. Let \(T_{L}\) contain the syntactically distinct formulae over one fixed letter \(p\) that admit proofs in which every formula, including axiom instances, contains at most \(L\) syntax-tree vertices. Size counts syntax-tree vertices, and the calculus contains the schemata \(K:A \rightarrow (B \rightarrow A)\) and \(S:\lbrack A \rightarrow (B \rightarrow C)\rbrack \rightarrow \lbrack(A \rightarrow B) \rightarrow (A \rightarrow C)\rbrack\), together with modus ponens. The resulting bound concerns explicit formula instances under these conventions.

\Needspace{5\baselineskip}
\textbf{Theorem 1.} For every integer \(m \geq 0\),

\[\fitdisplay{|T_{18m + 17}| \geq {Cat}_{m},\quad\quad{Cat}_{m} = \frac{1}{m + 1}\binom{2m}{m} \sim \frac{4^{m}}{\sqrt{\pi}\, m^{3/2}}.}\]

These conclusions contain \(4m + 3\) syntax-tree vertices and admit proofs with at most two applications of modus ponens along any path from an axiom instance. Appendix A presents the full proof and intermediate size calculations, using standard Catalan enumeration \citep{FlajoletSedgewick2009}.

An exhaustive store that retains each such formula separately requires exponentially many records in \(L\). No fixed total capacity can accommodate that representation for every increasing \(L\). This output-volume bound does not supply a new lower bound on the complexity of finding a single proof, and it permits streaming output, growing storage, and exhaustive computation of any fixed finite slice.

The generator needs a separate schema-level bound. Every \(F \rightarrow F\) instantiates the single schema \(A \rightarrow A\), so the explicit family behind Theorem 1 collapses under schema storage. Its exponential bound therefore \textbf{does not transfer} to the number of canonical schemata in the archive.

\textbf{Theorem 2.} Let \(H_{d}^{sch}\) contain \(K\) and every most general result of schematic modus ponens up to depth \(d\), with separate renaming of premise variables and canonicalisation only up to renaming. With no size or width limits, subsumption, or pruning,

\[\fitdisplay{|V(H_{d}^{sch})| \geq d + 1.}\]

Appendix A proves the claim. Together, the bounds establish why an ever-growing exact graph eventually exceeds fixed retained capacity, distinguishing explicit instances from canonical schemata. They motivate a design that limits active work and preserves proof certificates separately. Appendix A details the assumptions, pair counts, and finite restrictions.

\section*{5 Continuous skeleton and proof library}\label{continuous-skeleton-and-proof-library}
\addcontentsline{toc}{section}{5 Continuous skeleton and proof library}

\subsection*{5.1 Continuous growth with distinct states}\label{continuous-growth-with-distinct-states}
\addcontentsline{toc}{subsection}{5.1 Continuous growth with distinct states}

The continuous variant keeps one generator across epochs and reserves no compulsory time share for a second branch. We track its state as

\[\fitdisplay{X_{e} = (A_{e},P_{e},D_{e},\Omega_{e},\mathcal{I}_{e},B_{e}),}\]

where \(A_{e}\) contains active premises, \(P_{e}\) holds the passive queue of proven candidates, \(\Omega_{e}\) contains the certificate archive, \(\mathcal{I}_{e}\) contains working indices, \(B_{e}\) denotes remaining resources, and \(D_{e}\) contains root identifiers for closed library fragments whose certificates reside in \(\Omega_{e}\). In the continuation experiments the program defines retained anchors as \(S_{e} = A_{e} \cup D_{e}\), without an additional independent state component. The two sets serve distinct purposes: \(A_{e}\) supplies search premises, whereas \(D_{e}\) stores references to multistep proofs that the checker verified. Their ancestor proof support \(Anc(A_{e} \cup D_{e})\) serves checking and expansion, not an additional active base.

The archive grows through new canonical statements and their first proofs. The policy permanently protects only the initial axioms in the working set. The policy retains other active formulae for finite periods and extends those periods when a formula contributes to the first proof of a new result; library-selected formulae receive a separate finite period. Deactivation removes a formula from search indices but preserves its archive entry. If search rediscovers that formula, the policy can return it to the passive queue after a specified delay and within a return quota, measuring its new age from queue re-entry rather than old proof depth. These operations neither change the certificate nor turn a derived lemma into an axiom.

Explicit limits constrain active derived formulae and passive candidates after every layer. The policy first preserves the new search frontier, then accounts for finite library-retention periods and recent productive use. The policy sets per-layer selection width separately from total active-set size. Since the archive retains every entry, bounded activity does not imply constant total memory.

\subsection*{5.2 Utility of a reusable proof fragment}\label{utility-of-a-reusable-proof-fragment}
\addcontentsline{toc}{subsection}{5.2 Utility of a reusable proof fragment}

A closed fragment represents a particular first proof of a lemma with initial axioms at its leaves. A closed fragment excludes unproven open premises but may contain schema variables. To expand it, the checker follows each internal vertex's archived premises in their original order. This coding stage names existing proofs. The separate structural compiler below generalises short open proof programmes and derives their closed implication schemata. Both routes retain elementary certificates and leave the original consequence relation unchanged.

The continuation and network experiments in Sections 7.2--7.4 use the following proof-forest codec. The external comparison in Section 7.1 selects anchors by the reuse criterion in Section 6.3. For the codec, let \(R_e\) denote a finite set of recent proofs and \(pa(v)\) the ordered parent list, allowing repeated entries. Counting each premise occurrence gives the expanded proof-term size

\[\fitdisplay{t(v) = 1 + \sum_{p \in pa(v)}^{}t(p).}\]

The unit counts an axiom name, a rule application, or a fragment reference. For a chosen library \(D\), define

\[\fitdisplay{c_{D}(v) = \left\{ \begin{matrix}
1, & v \in D, \\
1 + \sum_{p \in pa(v)}^{}c_{D}(p), & v \notin D,
\end{matrix} \right.\ }\]

\[\fitdisplay{d_{D}(v) = 2 + \sum_{p \in pa(v)}^{}c_{D}(p).}\]

The quantity \(d_{D}(v)\) includes a definition header and its root instruction, whereas references to other definitions must follow the original well-founded dependencies. Thus the encoding gain for the proof forest accounts for both references and definitions:

\[\fitdisplay{G(D;R_{e}) = \sum_{v \in R_{e}}^{}t(v) - \left\lbrack \sum_{v \in R_{e}}^{}c_{D}(v) + \sum_{v \in D}^{}d_{D}(v) \right\rbrack.}\]

The proof-forest codec admits fragments only while they reduce code length; it need not fill its quota. Candidates comprise multistep proofs of height at least two that occur at least twice in the expanded forest; a bounded pool ranks them by how much each could reduce repetition. The greedy selector then adds the fragment with the largest positive increment in \(G\) and recalculates overlaps. The selector stops when no positive increment remains, so the procedure searches only a bounded candidate pool and does not guarantee optimal compression.

\textbf{Lemma 1 (library expansion).} For a finite acyclic archive, this encoding expands into the original forest of first proofs. If the selector starts from an empty library and admits only strictly positive increments in \(G\), its final gain on the training forest remains non-negative. Appendix B.5 proves the lemma.

The codec therefore recovers the original proofs without increasing the code length of the forest it uses to choose fragments. The gain \(G\) measures savings in expanded proof-term symbols; the archive already shares identical subproofs. The continuation experiments separately measure whether library conclusions improve future search under a CPU budget.

\subsection*{5.3 Continuous construction cycle}\label{continuous-construction-cycle}
\addcontentsline{toc}{subsection}{5.3 Continuous construction cycle}

A generation denotes a layer of new consequences, whereas an epoch marks when the program updates search state. Epoch number does not equal minimum proof height from the initial axioms. Unlike earlier restarts from selected anchors, this algorithm preserves one generator and its unfinished work; the experimental protocol specifies the retention periods, quotas, and bounds that govern that continuity.

\textbf{Algorithm 1. Continuous certified growth.}

\begin{enumerate}
\def\labelenumi{\arabic{enumi}.}
\tightlist
\item
  Check the initial axioms or inherited certified library, create the working indices, and initialise the archive and reusable dictionary.
\item
  While resources permit another layer, select an admissible frontier from the queue according to the fixed policy and its tie-breaking rule.
\item
  Process available rule applications. Check substitutions and side conditions, and canonicalise conclusions. Store a first proof for each new formula; handle rediscovery separately without increasing the new-formula count.
\item
  Update productive-use records and finite retention periods. Apply active-set and queue bounds while preserving certificates for inactive formulae. Return eligible rediscoveries to the queue subject to the specified delay and quota.
\item
  At each library review, select working anchors by the criterion for that experiment and give their conclusions time-limited premise slots. For the native run in Section 6.3, accumulate proof coverage and direct premise uses across the archive at each review. Keep its certificate after its slot expires.
\item
  Store the new certificate increment and state snapshot. Record the stopping reason. For the continuation experiments, credit completed layers. For the comparison in Section 6.3, keep every complete proof record emitted before growth stops.
\end{enumerate}

At a library-review point, the structural extension may add checked lemma schemata before the upper codec runs. It preserves the preceding interface and charges new definitions; it does not replace the accumulated generator with a new search from a hand-picked theorem list. 

The growth budget includes online anchor selection, index updates, internal checking, and storage. Section 6.3 gives both generators a separate allowance for final anchor selection. The independent checker replays the complete log after the experiment; we account for its cost separately. The pseudocode consequently specifies dependency order among operations, not equal execution times for successive layers.

\subsection*{5.4 Soundness and the limit of the guarantee}\label{soundness-and-the-limit-of-the-guarantee}
\addcontentsline{toc}{subsection}{5.4 Soundness and the limit of the guarantee}

Every archive certificate expands to the initial base provided the axioms satisfy the configuration, each new step passes checking, and each reference points to an earlier certified vertex. Appendices B.1--B.2 state the formal guarantees and prove them. These guarantees preserve existing proofs when the policy deactivates premises, but do not ensure complete enumeration: an inactive premise can prevent a useful future rule application.

Adding a provable lemma does not strengthen the underlying logic, but can change which proofs the available resources can reach \citep{CookReckhow1979}. Under size and height bounds, we must account explicitly for intermediate formulae when we expand proofs; Appendix B and Section C.1 provide the full resource conditions and a specific identity example. Complete finite saturation additionally requires the process to execute every available rule application eventually; the present bounded policy does not guarantee that schedule.

\subsection*{5.5 Learned pair selection with deferred replay}\label{learned-pair-selection-with-deferred-replay}
\addcontentsline{toc}{subsection}{5.5 Learned pair selection with deferred replay}

The later variant filters pairs of candidate premises before it unifies them. A frozen model scores only premise syntax and existing first-proof heights; it never inspects a proposed conclusion. Its training labels mark first-proof ancestry that supports an earlier compression dictionary. The policy retains roughly half of compatible candidates, includes a fixed exploratory share, and preserves their production traversal order. Thus it predicts a corridor towards useful proof support rather than the truth of an unseen theorem.

Deferred replay retains some skipped pairs in a bounded queue. In the continuation experiments, each epoch can test up to 128 ready pairs and reserve up to four ordinary premise slots to restore missing parents. A replay uses only currently active, certified premises; it does not introduce a new axiom or activate cold proof support. Queue capacity, overflow, and the cost of scores remain explicit. This mechanism delays the point at which the hard filter saturates while preserving the same elementary proof checker.

The policy constructs an archive, then extracts proof-backed anchor packages for neutral continuation. The reported chain CPU includes work to build the source, choose packages, and continue search. Training and the later independent audit enter separate accounts. Code-length gain chooses candidate support, and continued growth measures operational usefulness. These separate tests allow a compact package to succeed without claiming that every retained formula belongs to a unique mathematical core.

\subsection*{5.6 Additional compression above the accumulated library}\label{additional-compression-above-the-accumulated-library}
\addcontentsline{toc}{subsection}{5.6 Additional compression above the accumulated library}

The upper codec preserves a base dictionary \(D_{0}\) and adds a definition only when its joint marginal gain remains positive. It recalculates nested references and definition costs whenever it adds a definition and imposes no target dictionary size. For an eligible fragment, let \(r\) count its expanded occurrences in the current forest and dictionary bodies, and let \(b\) count the symbols in its current body. The exact saving under this code equals

\[\fitdisplay{\Delta(v \mid D) = (r - 1)(b - 1) - 2.}\]

The two charged symbols account for a new definition's root and header after the reference savings. Appendix F.1 proves this identity and its telescoping guarantee. Positive increments improve the previous encoding, even when \(D_{0}\) already contains mandatory entries; they do not guarantee a globally shortest code, fewer archive bytes, or faster future inference.

\subsection*{5.7 Structural synthesis of general lemma schemata}\label{structural-synthesis-of-general-lemma-schemata}
\addcontentsline{toc}{subsection}{5.7 Structural synthesis of general lemma schemata}

The structural layer abstracts a short MP proof tree, replaces its boundary by hypotheses, and considers possible equalities among those hypotheses. Joint typed unification solves all MP constraints together. For each consistent programme and order of its distinct boundary hypotheses, the compiler applies the deduction theorem and derives a closed implication schema. The generalisation rule (Gen) cuts an open fragment: the compiler never moves a deduction step through unrestricted Gen.

For example, a programme with boundary \(p\), \(p \rightarrow q\), and \(q \rightarrow r\) concludes \(r\). An appropriate discharge order yields

\[\fitdisplay{(p \rightarrow q) \rightarrow \left( (q \rightarrow r) \rightarrow (p \rightarrow r) \right).}\]

The compiler recovers this familiar composition law without a human theorem target. It expands every new step through ax-1, ax-2, and ax-mp, or invokes an existing checked schema through an explicit admissible instance. All output certificates pass the native checker and the independent Metamath verifier. Proposition F.2 states the proof obligation; this compiler implements the standard deduction construction rather than a new deduction theorem.

The catalogue distinguishes a requested principal formula from the actual schema that its certificate supplies. A more general stored schema may cover several requests, so we deduplicate delivered statements and their applicability frames. This distinction explains why the number of candidates need not match the number of rules that the compiler delivers. The finite grammar and size guards bound the actual construction, not all possible useful lemmas.

\subsection*{5.8 Fixed structural repertoire}\label{fixed-structural-repertoire}
\addcontentsline{toc}{subsection}{5.8 Fixed structural repertoire}

In a development comparison on the 919-statement reference corpus (Section 6.2), structural synthesis covered an average of 37.75 statements, compared with 33.58 for continued cumulative growth with proof coding. Both means exclude statements already covered by axiom instances and average over six source states and two neutral continuers after eight CPU-seconds of growth. A fixed grammar with at most two modus ponens nodes reproduced the same 21 requested formulae and identical certificates in all six source states. This result motivated our use of a fixed structural repertoire.

The native continuation experiments combine the cumulative constructor and positive-marginal codec with a finite structural repertoire. A common grammar delivers 217 checked schemata through three open MP nodes, including 16 short outputs through two nodes. The fixed quota-matched variant adds 12 extended outputs that the fixed ranking selects using pooled earlier development profiles. It uses no held-out theorem to select those rules. This fixed repertoire supplies a reproducible working configuration, not an optimum over all possible libraries.

The locked comparison separately tests whether the source history should change those 12 outputs, versus fixed, random, and whole-catalogue alternatives. The adaptive score belongs to that comparative experiment rather than the recommended architecture. Appendix F.5 specifies it fully, and the results retain its registered outcome.

\subsection*{5.9 Search permissions after the library freezes}\label{search-permissions-after-the-library-freezes}
\addcontentsline{toc}{subsection}{5.9 Search permissions after the library freezes}

The neutral test receives only declared interface statements as premises. It cannot use certificate-only support for sampling, subsumption, or rejecting a rediscovery as already known. If search derives a hidden support statement again, the checker first validates a new witness from currently permitted parents; only then does the statement gain search rights. A separate log retains that witness even when the old archive already stores the same statement.

The neutral continuers reset learned scores, deferred queues, and reuse counters. They apply either uniform premise sampling or the same fixed mixture of uniform, bridging, contracting, and frontier sampling to every arm. They retain admitted future results without the constructor's rot-pruning or active cap. Thus this test assesses a library's use under a common engine, not the effect of running each builder's controller longer. The mixed engine favours frontier statements according to their identifiers, so the experiment concerns delivered representations rather than an invariant semantic functional.

\section*{6 Materials and methods}\label{materials-and-methods}
\addcontentsline{toc}{section}{6 Materials and methods}

\subsection*{6.1 Formal systems and the scope of the tests}\label{formal-systems-and-the-scope-of-the-tests}
\addcontentsline{toc}{subsection}{6.1 Formal systems and the scope of the tests}

The propositional experiments use \(\rightarrow ,\neg\) with three axiom schemata:

\[\fitdisplay{\begin{matrix}
{ax}_{1}:A \rightarrow (B \rightarrow A), \\
{ax}_{2}:\lbrack A \rightarrow (B \rightarrow C)\rbrack \rightarrow \lbrack(A \rightarrow B) \rightarrow (A \rightarrow C)\rbrack, \\
{ax}_{3}:(\neg A \rightarrow \neg B) \rightarrow (B \rightarrow A),
\end{matrix}}\]

Schematic modus ponens renames premise variables apart before it unifies them. The modal condition adds the normal K axiom \(\square(A \rightarrow B) \rightarrow (\square A \rightarrow \square B)\) and necessitation for theorems. These two configurations differ in language, axioms, and rules together, so their contrast does not isolate a causal effect of logic alone.

In the propositional and modal continuation experiments, the bound \(L \in \{ 25,200\}\) counts abstract-syntax-tree nodes in each stored canonical formula. 

An independent checker traces each new result to the original axioms. It implements its own canonical forms, occurs-check unification, and rule checks without importing the generator's corresponding code. The audit validates one first proof per vertex, not the complete set of alternative proofs.

Each random initial scheduler state defines one trajectory and hence one repeat. In the continuation experiments, a common source archive defines a block within which we pair the ways of choosing anchors. Thousands of formulae along one trajectory do not supply thousands of independent observations.

\subsection*{6.2 Native development and the one-shot holdout}\label{native-development-and-the-one-shot-holdout}
\addcontentsline{toc}{subsection}{6.2 Native development and the one-shot holdout}

The native holdout and network experiments inherit a certified 2,708-statement library on the explicit basis that the network study specifies. They continue the cumulative constructor rather than rebuild that source from axioms. The constructor retains its learned half-filter on supported implication/negation terms, intrinsic fallback scores on other native syntax, deferred opportunities, four ordinary fragment slots, and a cap of 128 active derived premises. Rot-pruning deactivates an unleased, never-used premise after eight idle blocks; cap enforcement can deactivate other premises as well. Neither operation deletes its certificate.

Our earlier development reference contains 919 unique compatible closed native statements; axiom instances already cover 34. This reference informed method development. We use it to identify overlap with the held-out corpus.

For the one-shot test, Wernhard's POI supplies an external benchmark of 1,374 implication-form propositional theorems from Metamath \citep{Wernhard2026}. We exclude 402 alpha-exact matches to the preceding reference after unfolding the specified conjunction, disjunction, and equivalence definitions. Nine files retain named truth or falsehood constants outside the declared implication/negation language; we exclude rather than reinterpret them. The resulting corpus contains 963 unique targets. This filtering removes specified syntactic overlap, not all logical equivalence or one-way subsumption, and both references share Metamath ancestry.

The protocol fixes 16 new states of the random number generator (RNG) for source construction, five arms, and two neutral continuers, giving 160 trajectories but only 16 paired statistical blocks. Each common source runs 1,500 further opportunities. Arms receive the common compressed source and short catalogue, then add no extended rule, a fixed set of 12, a random set of 12, a source-adaptive set of 12, or all 201 extended entries. The full-catalogue control tests whether adaptation improves on supplying the entire finite structural repertoire rather than merely more rules.

The protocol requires all blind construction and continuation to finish and commit file hashes before the evaluator opens the targets once. No human target enters generation, features, admission, or the reward. The primary endpoint averages \(Q_{j}(m,60)\) across the two continuers within each source block. A second endpoint imposes 90 CPU-seconds of measured cold operational cost, charging new source work, the whole catalogue, arm preparation, and continuation setup before growth. Historical costs before the inherited checkpoint, interpreter startup, offline goal observation, terminal bulk export, and independent audit remain separate; this endpoint does not measure a complete from-axioms prover job.

We compare adaptive selection with fixed, random, and full-catalogue controls through exact one-sided paired sign flips, assuming sign exchangeability under the null. For each comparison, we retain the larger p-value from the two endpoints, then apply Holm's correction across the three controls. Confirmation requires an adjusted p below 0.05 and a mean gain of at least one residual target at both endpoints against every control. All planned cells and CPU horizons must finish, the selector must vary across sources, and the independent certificate and permission audit must pass. Missing cells, early guards, or a constant selector preclude confirmation. Appendix G records the frozen commitment and outcome.

The native bound of 160 constrains stored canonical formulae, and the certificate-height bound reaches 256. These bounds do not constrain every fully instantiated intermediate formula after expanding all substitutions, unlike the explicit-instance convention in Theorem 1. Native trajectories and the earlier propositional/modal panels therefore retain separate labels, metrics, and cost accounts.

\subsection*{6.3 Freezing a skeleton for external proofs}\label{external-methods}
\addcontentsline{toc}{subsection}{6.3 Freezing a skeleton for external proofs}

Each generator starts with ax-1, ax-2, and ax-3 from Section 6.1. Both use only condensed detachment. They rename variables apart, unify the minor premise with the antecedent of the major premise, and apply modus ponens. Our constructor chooses premise pairs from its forward search. SGCD uses Wernhard's single-worker bottom-up enumeration (\texttt{inc\_tsize1}, duplicate filtering, and a level trim of 1,100). Each run starts cold on one pinned core with 1~GiB of memory and a growth ceiling of 90 CPU-seconds. We repeat the comparison at a 15-CPU-second ceiling. Both generators use the same export procedure with a further ceiling of 90 CPU-seconds per archive. In the three 90-second controls, SGCD completed its enumeration in 32.26--37.38 CPU-seconds.

To select the final anchors, we keep the first checked proof of each distinct schema in both archives and consider candidates of proof height at least two. A candidate earns one vote for each still-uncovered retained proof that contains it as a proper ancestor. To qualify on coverage alone, it must occur in at least two distinct retained proofs.

At export, we also count the distinct inferences in the retained archive for which a candidate could supply a multistep premise. One substitution must turn the candidate schema into the whole premise and agree at every occurrence of a repeated variable. The propositional comparison uses a single sort. In the typed implementation, substitutions preserve sorts, and schemata with distinct-variable conditions receive credit only for recorded uses as a premise. An inference counts once per candidate even if both premises match. Only candidates on the Pareto frontier for smaller formula size and more such inferences receive these votes. The votes remain available even after another anchor covers the same proof roots. We choose the candidate with the largest sum of coverage and inference votes. Ties favour greater additional coverage, then lower proof height, then earlier archive order. We stop when no candidate earns a positive score. This rule determines the final bank size. Neither generator sees the external proofs.

We designate native seed 101 as the primary run and report four further seeds (202, 303, 404, and 505). Three runs check the repeatability of deterministic SGCD. We score a frozen bank by replacing proper subproofs, as in Section 3.4, and replay the shortened proofs. Our first proof corpus comes from the \emph{Aviary} catalogue of typed combinators \citep{TetleyAviary}. We translate explicit function bodies by standard \(S,K\) bracket abstraction with \(\eta\)-contraction and check the resulting theorem against its declared type. The second source comprises seven published SGCD runs \citep{Wernhard2026}. Appendix I details the filters and identifies which proofs and statements are new.

\section*{7 Results}\label{results}
\addcontentsline{toc}{section}{7 Results}

\subsection*{7.1 The ten-anchor skeleton on external proofs}\label{external-skeleton}
\addcontentsline{toc}{subsection}{7.1 The ten-anchor skeleton on external proofs}

The primary run retained ten anchor schemata from 17,202 distinct derived statements. Their certificates share 27 applications of condensed detachment. No inference in this run cites the node for \(A\to A\) directly. Yet 39 distinct inferences use a premise that this schema can supply by substitution. Those uses give the identity fifth place in the primary bank. All five native seeds retain it at both growth ceilings. Appendix I gives the ten primary anchors and their complete certificate. The independent Metamath verifier accepted every exported anchor.

On the full Aviary panel (41 types), the primary skeleton shortens 38 proofs and saves 104 steps. SGCD shortens 39 and saves 146. Table~\ref{tab:external-skeleton} gives the results after filtering for new proof structures or statements. On the 30 previously unseen Aviary proof structures, our ten anchors and SGCD's 22 anchors both shorten every proof. This gives the primary skeleton a smaller full bank at equal coverage. SGCD saves more steps in this run. If we truncate SGCD's bank to its first ten anchors, it still shortens all 30 and saves 115 steps, compared with our 90. The difference in full bank size therefore describes what the shared selection rule returns.

\begin{table}[htbp]
\centering
\caption{Inference steps saved by frozen skeletons at the 90-CPU-second growth ceiling. The native column gives the result for seed 101 and the range across all five seeds.}\label{tab:external-skeleton}
\small\setlength{\tabcolsep}{4pt}
\begin{tabular}{@{}lrrr@{}}
\toprule
External proofs & Original steps & Native: primary [range] & SGCD\\
\midrule
Aviary, 30 new proofs & 412 & 90 [90--141] & 128\\
Aviary, 4 new statements & 76 & 9 [9--16] & 15\\
SGCD, 475 new proofs & 10,898 & 1,180 [856--1,636] & 1,308\\
\bottomrule
\end{tabular}
\end{table}

After charging the shared certificate once per bank, the primary native and SGCD banks save 63 and 100 steps, respectively, on the 30 Aviary proofs. On the four previously unseen Aviary statements, both primary banks shorten all four proofs. Four statements provide little evidence about transfer to new theorem statements.

On the 475 previously unseen SGCD proofs of familiar theorems, the primary native bank shortens 411 proofs and the SGCD bank shortens all 475. Appendix I.2 gives the results for each seed. At the 15-CPU-second growth ceiling, the five native banks save 87--134 steps on the 30 Aviary proofs and 727--1,308 on the 475 published proofs. SGCD saves 128 and 1,308 steps on those panels, respectively. These counts measure how much of a supplied derivation an anchor can replace.

\subsection*{7.2 Neutral continuation and empirical Pareto trade-offs}\label{neutral-continuation-and-empirical-pareto-trade-offs}
\addcontentsline{toc}{subsection}{7.2 Neutral continuation and empirical Pareto trade-offs}

The independent continuation panel comprises 144 source trajectories, six pair-selection policies, three logic--size regimes, and eight scheduler states per regime. Each source supplies packages with up to 4, 16, or 64 anchors at nominal source ceilings of 60, 180, and 600 CPU-seconds. Separate neutral generators continue from those packages, giving 1,296 continuations with nested readings at 10, 30, and 60 CPU-seconds. The primary endpoint fixes the source ceiling at 180 CPU-seconds, the anchor cap at 16, and continuation at 30 CPU-seconds. Novelty excludes the union of concurrent source archives and the frozen training archives; Appendix D records the protocol and checks.

At this endpoint, the learned half-filter with deferred replay improves mean novelty over ungated growth in both propositional regimes, with eight positive pairs out of eight. The size-25 regime also improves the measured CPU and package-byte coordinates in every pair. The size-200 package grows by 16.5\% on average, illustrating why a useful skeleton need not minimise every resource coordinate. The modal comparison splits four wins and four losses despite a positive mean (Table 2).

\begingroup
\small
\setlength{\tabcolsep}{3pt}
\renewcommand{\arraystretch}{1.13}

\begin{longtable}[]{@{}
  >{\raggedright\arraybackslash}p{(\linewidth - 8\tabcolsep) * \real{0.1923}}
  >{\raggedleft\arraybackslash}p{(\linewidth - 8\tabcolsep) * \real{0.2053}}
  >{\raggedleft\arraybackslash}p{(\linewidth - 8\tabcolsep) * \real{0.1989}}
  >{\raggedleft\arraybackslash}p{(\linewidth - 8\tabcolsep) * \real{0.1307}}
  >{\raggedleft\arraybackslash}p{(\linewidth - 8\tabcolsep) * \real{0.2729}}@{}}
\caption{Neutral-continuation outcomes for ungated growth and learned half-selection with replay}\label{tab:3}\tabularnewline
\toprule\noalign{}
\begin{minipage}[b]{\linewidth}\raggedright
\textbf{Regime}
\end{minipage} & \begin{minipage}[b]{\linewidth}\raggedleft
\textbf{New schemata without gate}
\end{minipage} & \begin{minipage}[b]{\linewidth}\raggedleft
\textbf{Learned half with replay}
\end{minipage} & \begin{minipage}[b]{\linewidth}\raggedleft
\textbf{Positive pairs}
\end{minipage} & \begin{minipage}[b]{\linewidth}\raggedleft
\textbf{Chain CPU without gate / with replay}
\end{minipage} \\
\midrule\noalign{}
\endfirsthead
\toprule\noalign{}
\begin{minipage}[b]{\linewidth}\raggedright
\textbf{Regime}
\end{minipage} & \begin{minipage}[b]{\linewidth}\raggedleft
\textbf{New schemata without gate}
\end{minipage} & \begin{minipage}[b]{\linewidth}\raggedleft
\textbf{Learned half with replay}
\end{minipage} & \begin{minipage}[b]{\linewidth}\raggedleft
\textbf{Positive pairs}
\end{minipage} & \begin{minipage}[b]{\linewidth}\raggedleft
\textbf{Chain CPU without gate / with replay}
\end{minipage} \\
\midrule\noalign{}
\endhead
\bottomrule\noalign{}
\endlastfoot
Propositional, size 25 & 13,935.38 & 27,285.12 & 8/8 & 209.74 / 169.53 \\
Propositional, size 200 & 20,569.25 & 26,805.12 & 8/8 & 147.54 / 136.06 \\
Modal K, size 25 & 22,164.25 & 25,456.50 & 4/8 & 209.68 / 139.00 \\
\end{longtable}

\endgroup

The table reports means across eight scheduler states, not independent mathematical theories. Chain CPU includes credited work to build sources, select packages, and continue search, whereas training and later independent audit remain separate. Resource guards stop 119 sources at the memory threshold, 21 through the implementation's saturation condition, and four at the epoch cap; no source completes 600 CPU-seconds. The gains therefore characterise the observed resource-bounded trajectories rather than a completed long-horizon limit or a fixed-quality speedup.

The Pareto comparison fixes the primary source ceiling and continuation horizon while varying six policies and three anchor caps. Each regime contributes 18 mean triples. A setting enters the three-coordinate frontier when no other setting uses no more CPU or package bytes, yields at least as much novelty, and strictly improves at least one coordinate. The figure projects those triples onto CPU and novelty; outlined points mark the full three-coordinate frontier, so a point can remain nondominated through smaller package size even when its projection looks inferior (Fig. 1).

\begin{figure}[!htbp]
\centering
\includegraphics[width=\linewidth,height=.67\textheight,keepaspectratio]{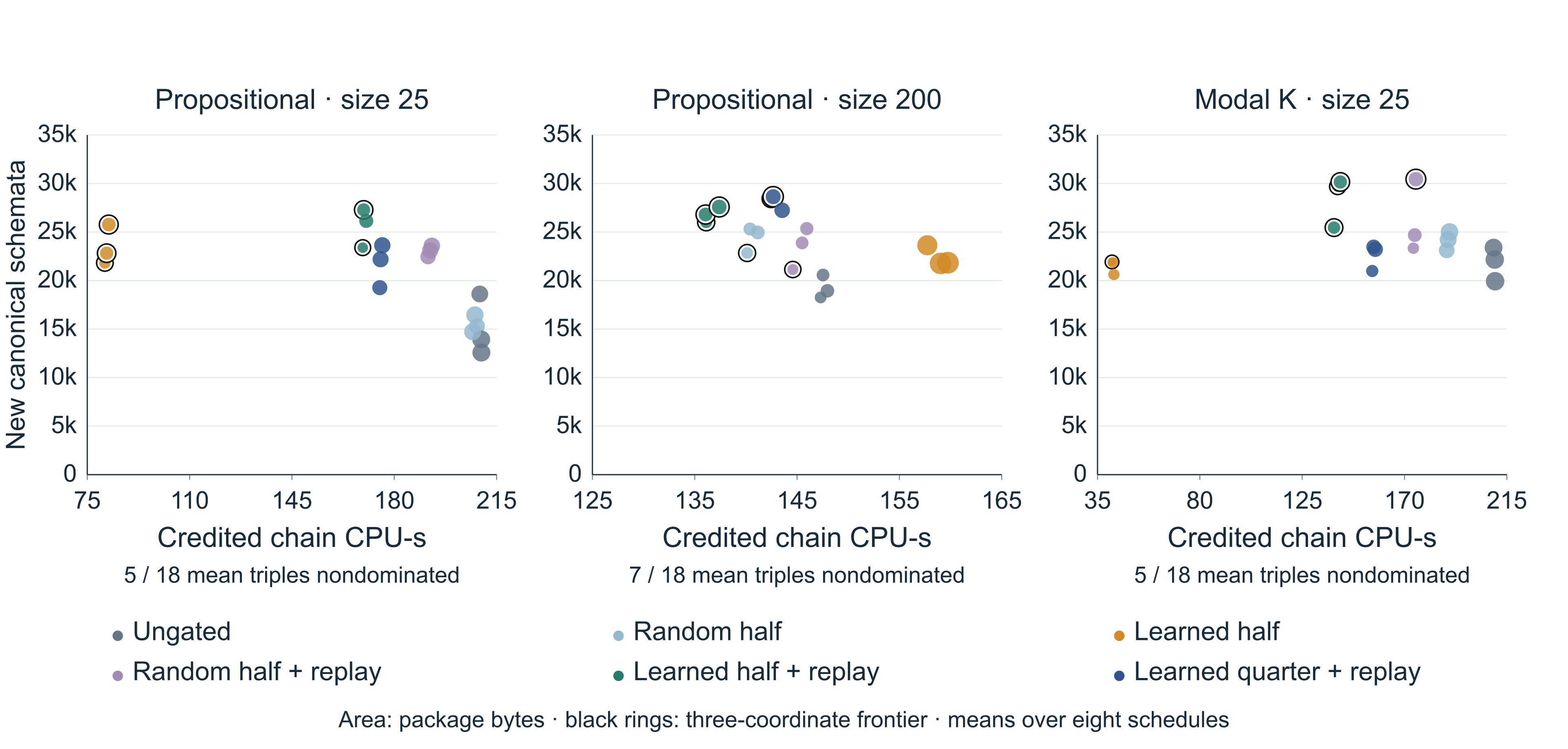}
\caption{Empirical Pareto trade-offs at a source ceiling of 180 CPU-seconds and continuation of 30 CPU-seconds. Colour identifies the pair policy, marker area indicates proof-backed package bytes, and outlines identify nondominated mean triples across CPU, bytes, and novelty. Each point averages eight scheduler states}\label{fig:2}
\end{figure}

These trade-offs favour different settings for different budgets. The learned half-filter without replay can cost less than replay at size 25, whereas replay yields more new schemata. More anchors or a later source slice need not improve continuation. The empirical frontier consequently supports a choice among useful finite skeletons, rather than a universal winner or an optimum beyond this grid.

\subsection*{7.3 Locked native comparison}\label{locked-native-comparison}
\addcontentsline{toc}{subsection}{7.3 Locked native comparison}

All 16 source blocks and 160 planned continuations finish both registered horizons and pass independent replay. For each table entry, we first average the two neutral continuers within each source, then average across the 16 sources. Initial coverage counts all 963 reference targets; both future columns count only the common initially uncovered residual within that source block. Derived-anchor counts exclude 29 basis roots and separate certificate-only support (Table 3).

\Needspace{19\baselineskip}
\begingroup
\small
\setlength{\tabcolsep}{3pt}
\renewcommand{\arraystretch}{1.13}

\begin{longtable}[]{@{}
  >{\raggedright\arraybackslash}p{(\linewidth - 8\tabcolsep) * \real{0.2558}}
  >{\raggedleft\arraybackslash}p{(\linewidth - 8\tabcolsep) * \real{0.1701}}
  >{\raggedleft\arraybackslash}p{(\linewidth - 8\tabcolsep) * \real{0.1865}}
  >{\raggedleft\arraybackslash}p{(\linewidth - 8\tabcolsep) * \real{0.2259}}
  >{\raggedleft\arraybackslash}p{(\linewidth - 8\tabcolsep) * \real{0.1617}}@{}}
\caption{Initial coverage, further common-residual coverage, and anchors in the locked native comparison}\label{tab:5}\tabularnewline
\toprule\noalign{}
\begin{minipage}[b]{\linewidth}\raggedright
\textbf{Library}
\end{minipage} & \begin{minipage}[b]{\linewidth}\raggedleft
\textbf{Initial coverage}
\end{minipage} & \begin{minipage}[b]{\linewidth}\raggedleft
\textbf{New at 60 growth s}
\end{minipage} & \begin{minipage}[b]{\linewidth}\raggedleft
\textbf{New at 90 operational s}
\end{minipage} & \begin{minipage}[b]{\linewidth}\raggedleft
\textbf{Derived anchors}
\end{minipage} \\
\midrule\noalign{}
\endfirsthead
\toprule\noalign{}
\begin{minipage}[b]{\linewidth}\raggedright
\textbf{Library}
\end{minipage} & \begin{minipage}[b]{\linewidth}\raggedleft
\textbf{Initial coverage}
\end{minipage} & \begin{minipage}[b]{\linewidth}\raggedleft
\textbf{New at 60 growth s}
\end{minipage} & \begin{minipage}[b]{\linewidth}\raggedleft
\textbf{New at 90 operational s}
\end{minipage} & \begin{minipage}[b]{\linewidth}\raggedleft
\textbf{Derived anchors}
\end{minipage} \\
\midrule\noalign{}
\endhead
\bottomrule\noalign{}
\endlastfoot
Common core and short rules & 25.00 & 1.312 & 1.344 & 633.6 \\
Fixed extended repertoire & 29.00 & 1.594 & 1.625 & 675.0 \\
Random extended repertoire & 26.38 & 1.500 & 1.562 & 680.4 \\
Adaptive test variant & 27.88 & 1.656 & 1.688 & 673.4 \\
Whole fixed catalogue & 44.00 & 1.562 & 1.594 & 1282.3 \\
\end{longtable}

\endgroup

The whole fixed catalogue initially covers 44 targets, versus 29 for the fixed extended repertoire, but reaches 1.56 versus 1.59 new common-residual targets at 60 growth CPU-seconds. This contrast illustrates why initial coverage cannot rank future capability. The adaptive score chooses three distinct rule sets and reaches 1.656 new residual targets, yet its paired advantage over the quota-matched fixed repertoire averages only 0.0625 targets at either endpoint (Table 4).

\begingroup
\small
\setlength{\tabcolsep}{3pt}
\renewcommand{\arraystretch}{1.13}

\begin{longtable}[]{@{}
  >{\raggedright\arraybackslash}p{(\linewidth - 6\tabcolsep) * \real{0.3000}}
  >{\raggedleft\arraybackslash}p{(\linewidth - 6\tabcolsep) * \real{0.2400}}
  >{\raggedleft\arraybackslash}p{(\linewidth - 6\tabcolsep) * \real{0.2400}}
  >{\raggedleft\arraybackslash}p{(\linewidth - 6\tabcolsep) * \real{0.2200}}@{}}
\caption{Paired adaptive-minus-control differences and Holm-adjusted tests at both registered horizons}\label{tab:6}\tabularnewline
\toprule\noalign{}
\begin{minipage}[b]{\linewidth}\raggedright
\textbf{Adaptive minus control}
\end{minipage} & \begin{minipage}[b]{\linewidth}\raggedleft
\textbf{Difference at 60 growth s}
\end{minipage} & \begin{minipage}[b]{\linewidth}\raggedleft
\textbf{Difference at 90 operational s}
\end{minipage} & \begin{minipage}[b]{\linewidth}\raggedleft
\textbf{Holm adjusted p}
\end{minipage} \\
\midrule\noalign{}
\endfirsthead
\toprule\noalign{}
\begin{minipage}[b]{\linewidth}\raggedright
\textbf{Adaptive minus control}
\end{minipage} & \begin{minipage}[b]{\linewidth}\raggedleft
\textbf{Difference at 60 growth s}
\end{minipage} & \begin{minipage}[b]{\linewidth}\raggedleft
\textbf{Difference at 90 operational s}
\end{minipage} & \begin{minipage}[b]{\linewidth}\raggedleft
\textbf{Holm adjusted p}
\end{minipage} \\
\midrule\noalign{}
\endhead
\bottomrule\noalign{}
\endlastfoot
fixed & +0.06250 & +0.06250 & 1.000 \\
random & +0.15625 & +0.12500 & 1.000 \\
fixed all & +0.09375 & +0.09375 & 1.000 \\
\end{longtable}

\endgroup

Every Holm-adjusted p-value equals 1.000, and no contrast reaches the registered mean margin of one additional target at both clocks. We therefore retain fixed structural rules in the working architecture and do not attribute a confirmed gain to adaptive catalogue ranking. Appendix G reports the full paired comparison, statistical rule, and technical amendment.

The independent audit reconstructs 4,174,478 admission events and 2,846 witnesses that record how search rediscovers hidden support, checks all target positions, and reproduces the endpoints and decision. These counts include repeated events across trajectories, not independent theorems. The comparison charges the full recorded catalogue to every arm under the shared backend; it does not test a specialised static implementation that avoids compiling unused entries.

\subsection*{7.4 Metamath and generated proof networks}\label{metamath-and-generated-proof-networks}
\addcontentsline{toc}{subsection}{7.4 Metamath and generated proof networks}

The network comparison follows the cumulative method through its delivered library and certificate support. Panel b uses the first source, 2026102201, with the quota-matched fixed structural repertoire. Appendix H presents the profiles of both variants and the separately preselected adaptive image. Both representatives follow the first-source convention, independently of target coverage, graph appearance, and power-law p-values.

All objects share the explicit native basis from the pinned \path{set.mm} revision \citep{MetamathSetMMPinned}: 25 logical or set-theoretic axioms and rules, six definitions, and 12 syntax assertions. Twenty-nine closed source schemata initialise search. The inherited proof-closed core contains 2,708 statements; the cumulative constructor continues that core, filters and revisits opportunities, applies reversible rot-pruning, compiles structural rules, and codes recurrent proofs. Panel b shows the delivered interface and every ancestor that its certificates require, not the full inherited archive. Appendix H records the sources, projections, and metrics.

The first-source fixed library provides 675 derived anchors and 29 basis roots; 137 further vertices supply certificate-only support. Its projection therefore contains 841 vertices and 1,573 directed arcs. Across all 16 fixed sources, anchor counts range from 665 to 682, with a mean of 675.0; projected sizes range from 836 to 852. Hidden support does not confer search rights, and a verifier can expand every anchor to the original basis.

We project the proof hypergraph onto an ordinary directed graph to measure paths, clusters, and communities. For each inference, we draw an arc from each premise to its conclusion, omit primitive rule vertices, and merge repeated arcs. The hypergraph and its certificates record which premises a rule requires together and what substitutions and side conditions it uses. The human fragment contains 2,306 named assertions whose actual proofs use only the declared basis, including assertions with essential hypotheses. Native vertices denote closed schemata with mandatory distinct-variable conditions. A shared basis does not eliminate the difference between named human steps and elementary native inferences.

The full reference enumerates all admitted MP steps through most-general unification and prescribed generalisations through proof height three, without skeleton-selection or formula-size cuts. Its 11,772 statements support both first-proof and all-inference projections. This reference supplies a complete finite slice of the stated operator, not all substitution instances, a deductive fixed point, or infinite closure of Zermelo--Fraenkel set theory with choice (ZFC). Panel b reaches certificate height 23; the comparison holds neither depth nor construction cost constant (Fig. 2).

\begin{figure}[!htbp]
\centering
\includegraphics[width=\linewidth,height=.67\textheight,keepaspectratio]{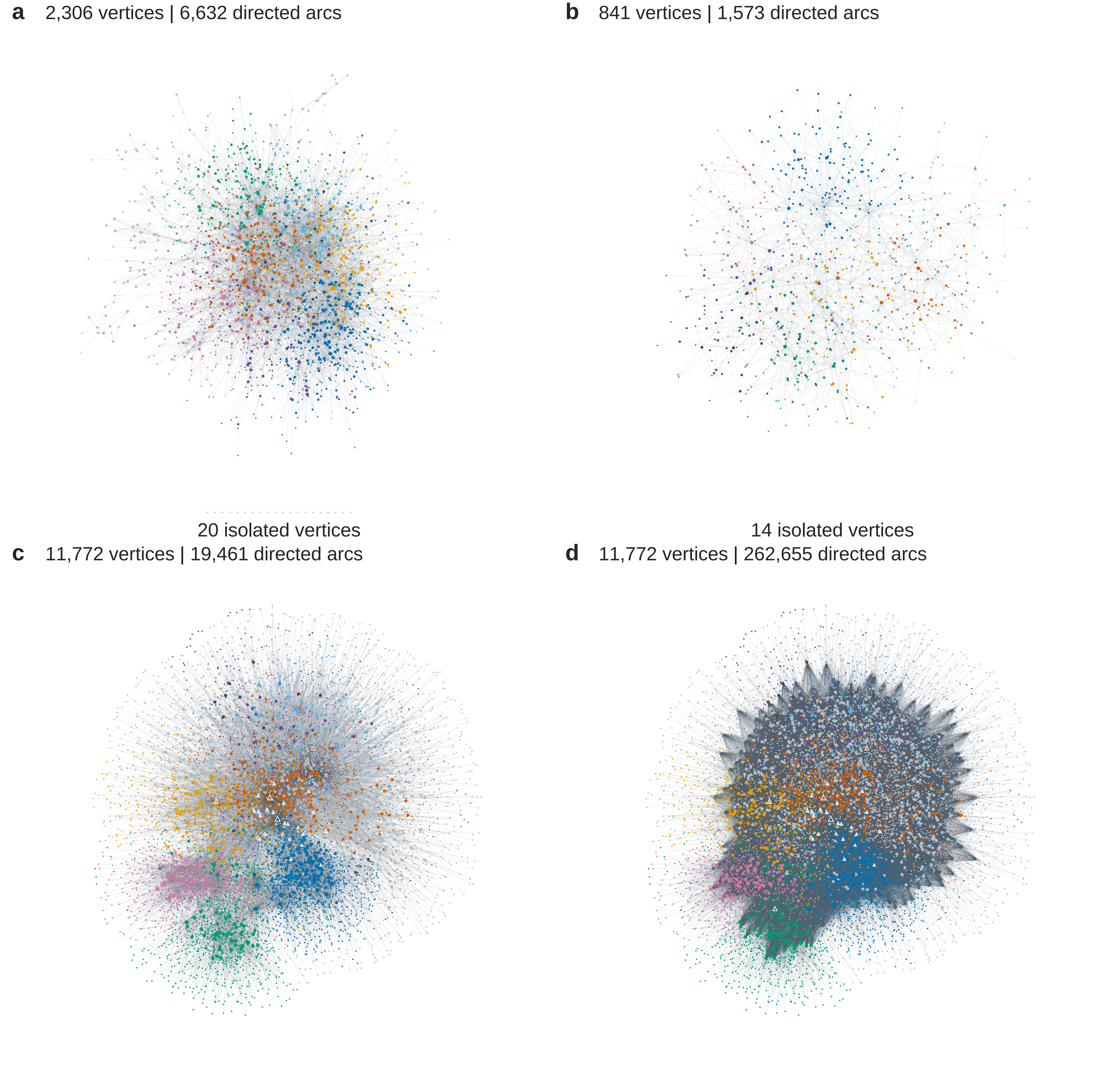}
\caption{Native proof networks on one explicit basis. a shows the compatible named Metamath fragment; b shows the first-source cumulative skeleton with fixed structural rules, including all certificate support; c shows first proofs in the complete height-three operator slice; d shows all enumerated inferences on those same vertices. Every vertex and unique undirected edge enters the drawing; only arrowheads disappear. Colours distinguish the eight largest communities in a, b, and c, with smaller communities in grey. d inherits c's coordinates, colours, and vertex areas; its tabulated modularity follows its own community search. Triangles mark the 29 basis roots. Areas increase with log reuse. Separate layouts do not measure comparable semantic distances}\label{fig:4}
\end{figure}

The representative skeleton reaches modularity 0.622, versus 0.539 across ten degree-preserving controls, and outgoing distinct-use degree yields a Gini coefficient of 0.731. Across the 16 fixed sources, modularity ranges from 0.622 to 0.629, with a mean of 0.625. These measurements characterise concentrated reuse and modular structure without promoting a single image to a mean result.

The outgoing-degree profiles also differ from the earlier 2,708-vertex core. None of the 16 fixed-repertoire sources rejects the pure discrete power-law family at five percent; bootstrap p-values range from 0.24190 to 0.62575. The representative fits an exponent of 2.044719 from degree one, with 524 tail observations and p = 0.54640. Its other 317 vertices carry zero outgoing degree. On this same tail, the cutoff power law and rounded lognormal yield descriptive log likelihoods of \ensuremath{-}815.753 and \ensuremath{-}816.024, versus \ensuremath{-}818.182 for the pure power law. Compatibility with the pure family therefore does not select it uniquely or establish an infinite scale-free growth mechanism (Table 5).

\begingroup
\small
\setlength{\tabcolsep}{3pt}
\renewcommand{\arraystretch}{1.13}

\begin{longtable}[]{@{}
  >{\raggedright\arraybackslash}p{(\linewidth - 8\tabcolsep) * \real{0.3199}}
  >{\raggedleft\arraybackslash}p{(\linewidth - 8\tabcolsep) * \real{0.1700}}
  >{\raggedleft\arraybackslash}p{(\linewidth - 8\tabcolsep) * \real{0.1700}}
  >{\raggedleft\arraybackslash}p{(\linewidth - 8\tabcolsep) * \real{0.1700}}
  >{\raggedleft\arraybackslash}p{(\linewidth - 8\tabcolsep) * \real{0.1700}}@{}}
\caption{Network metrics for Metamath, the cumulative skeleton, and two complete height-three projections}\label{tab:7}\tabularnewline
\toprule\noalign{}
\begin{minipage}[b]{\linewidth}\raggedright
\textbf{Metric}
\end{minipage} & \begin{minipage}[b]{\linewidth}\raggedleft
\textbf{a Metamath}
\end{minipage} & \begin{minipage}[b]{\linewidth}\raggedleft
\textbf{b Cumulative skeleton}
\end{minipage} & \begin{minipage}[b]{\linewidth}\raggedleft
\textbf{c First proofs}
\end{minipage} & \begin{minipage}[b]{\linewidth}\raggedleft
\textbf{d All inferences}
\end{minipage} \\
\midrule\noalign{}
\endfirsthead
\toprule\noalign{}
\begin{minipage}[b]{\linewidth}\raggedright
\textbf{Metric}
\end{minipage} & \begin{minipage}[b]{\linewidth}\raggedleft
\textbf{a Metamath}
\end{minipage} & \begin{minipage}[b]{\linewidth}\raggedleft
\textbf{b Cumulative skeleton}
\end{minipage} & \begin{minipage}[b]{\linewidth}\raggedleft
\textbf{c First proofs}
\end{minipage} & \begin{minipage}[b]{\linewidth}\raggedleft
\textbf{d All inferences}
\end{minipage} \\
\midrule\noalign{}
\endhead
\bottomrule\noalign{}
\endlastfoot
Vertices \(N\) & 2306 & 841 & 11772 & 11772 \\
Directed arcs \(E\) & 6632 & 1573 & 19461 & 262655 \\
Largest-component vertices & 2286 & 827 & 11772 & 11772 \\
Isolated vertices & 20 & 14 & 0 & 0 \\
Mean local clustering \(C\) & 0.0689 & 0.0202 & 0.0347 & 0.3382 \\
Mean shortest path \(L\) & 4.137 & 4.177 & 4.334 & 3.213 \\
Diameter \(D\) & 11 & 8 & 6 & 4 \\
Modularity \(Q\) & 0.568 & 0.622 & 0.675 & 0.058 \\
Reference modularity \(Q_{0}\) & 0.401 & 0.539 & 0.599 & 0.053 \\
Small-world ratio \(\sigma_{DP}\) & 3.990 & 0.636 & 0.592 & 0.417 \\
Tail observations & 205 & 524 & 63 & 41 \\
Pure power law p & 0.09895 & 0.54640 & 0.01080 & 0.36490 \\
\end{longtable}

\endgroup

Table columns follow the figure panels. C, L, D, and Q refer to simple undirected projections. C assigns zero below degree two. L and D refer to the largest connected component and use every observed breadth-first-search (BFS) source. Panel a contains 20 isolates, and panel b contains 14 unused basis roots. Panels c and d remain connected. Q0 averages ten simple degree-preserving controls, with three Louvain starts for every observed and control graph. Panel b's control paths also use every source; the retained c and d controls estimate paths from 64 sources. The tail test for outgoing distinct-use degree fixes 20,000 bootstrap draws and refits both threshold and exponent. Appendix H reports all 32 new profiles and their scope.

The c--d pair isolates an edge-recording effect on identical statements: alternative edges raise clustering from 0.0347 to 0.3382, shorten the mean path from 4.334 to 3.213, and lower modularity from 0.675 to 0.058. No logical or vertex change explains that contrast. The ordinary proof-network profile therefore depends materially on how the representation records alternative derivations.

Newman (\citeyear{Newman2003}), in \emph{The Structure and Function of Complex Networks}, treats degree heterogeneity, clustering, paths, and correlations as complementary features. Humphries and Gurney (\citeyear{HumphriesGurney2008}), in \emph{Network `Small-World-Ness': A Quantitative Method for Determining Canonical Network Equivalence}, formalise the clustering--path comparison behind the small-world ratio. Under the declared degree-preserving reference, Metamath reaches a ratio of 3.990, whereas the skeleton reaches 0.636. Thus, the skeleton exhibits concentrated reuse, modular organisation, and a compatible conditional power-law tail, but its short paths do not establish small-world structure. Each projection defines an ordinary graph. We describe it as a complex network when we study how its links concentrate, how its communities form, and how far apart its vertices lie. A conditional tail fit does not establish infinite scale invariance, and network metrics complement rather than replace future-use tests \citep{ClausetEtAl2009,BroidoClauset2019}. Ten null graphs support descriptive contrasts rather than a rank-test claim below five percent.

\section*{8 Discussion and scope}\label{discussion-and-scope}
\addcontentsline{toc}{section}{8 Discussion and scope}

Individual anchors contribute differently to transfer. In a post hoc check, we kept only the first anchor of the primary bank (node 6 in Appendix I.1). Every native seed and all three SGCD repeats retained that schema at the 90-CPU-second ceiling. Used alone, it saves 52 steps across 26 of the 30 Aviary proofs and 659 steps across 331 of the 475 published SGCD proofs. Nine of the ten primary schemata also occur in the full SGCD bank. On Aviary those nine alone save 80 steps across 29 of the 30 proofs. Adding the remaining schema (node 107), \(A\to(B\to(C\to A))\), brings the total to 90 steps across all 30. A separate test could hold the growth archive fixed and compare our rule with selection by use frequency and random sampling at the same bank size.

The external readout acts on each proof DAG and counts a shared inference once within that proof. The codec used in the continuation experiments acts on expanded proof terms. Its symbol savings and the external DAG step counts use different units. The 475-proof corpus also shares its proof-generation tradition with SGCD, while Aviary samples typed combinators in the implicational fragment. Neither corpus represents all of mathematics.

Human references expose differences that raw formula counts conceal. The locked comparison evaluates a previously unscored 963-target POI subset only after every search trajectory freezes. An independent audit then checks the outcome. These corpora neither supply a complete modal reference nor assess the full mathematical value of the generated library.

The native network study adds a structural result to the operational evidence. A proof-closed skeleton exhibits modular organisation and concentrated reuse, whereas the fixed-vertex reference shows that alternative proof edges can strongly alter the same network metrics. Verified certificates, human-statement coverage, and measured continuation complement this structural profile. The native network study and the earlier pair-selection trajectories retain their separate scopes.

The combinatorial claims also require distinct scopes. The exponential lower bound counts explicit theorem instances, whereas the result for canonical schemata establishes only unboundedness. Neither the theorems nor finite runs establish non-computability of a bounded slice, exponential growth of schemata, or the limiting topology of an infinite proof space. Since the two logics change their language, axioms, and rules together, they do not isolate logic alone as a cause.

The native comparison supplies a reproducible operating configuration with fixed structural rules. Goal-directed proof search and control towards prescribed graph properties define the next two research tracks. For knowledge-based AI, our method preserves certified knowledge, changes which statements inference may use, and accounts for the required resources.

\section*{9 Conclusions}\label{conclusions}
\addcontentsline{toc}{section}{9 Conclusions}

The finite semantic skeleton presented here consists of ten reusable statements and a shared graph of 27 inference steps. Every anchor reference expands to a proof from the three axioms. An external checker can reconstruct the skeleton from Appendix I alone. A reader can use any of the ten statements as a premise and expand its certificate when needed.

On the 30 new Aviary proof structures, the primary skeleton saves 90 of 412 inference steps. Its shared certificate costs 27 steps once for the whole panel, leaving a net saving of 63. The saving varies across proof corpora and across runs from the same axioms (Table~\ref{tab:external-skeleton}).

The continuation panels describe what further growth can reach from a certified library. The network study shows that, on the same 11,772 statements, recording alternative inferences changes clustering from 0.0347 to 0.3382 and modularity from 0.675 to 0.058. Both the choice of anchors and the choice of proof edges matter when interpreting a finite fragment.

\Needspace{18\baselineskip}
\subsection*{9.1 Extending the skeleton towards a distant goal in Lean}\label{applying-the-skeleton-in-lean}
\addcontentsline{toc}{subsection}{9.1 Extending the skeleton towards a distant goal in Lean}

A proof of a functional equation, for example, may require a decay bound for an auxiliary kernel. We propose to work backward from the theorem to such intermediate statements and grow the skeleton forward from the available premises.

\begin{enumerate}
\item \textbf{Choose an intermediate goal.} Give a large language model (LLM) the target theorem and the open goals in Lean. Ask it to suggest a mathematical area in which to seek a bridge from an open goal to known lemmas.
\item \textbf{Find the missing premises.} Try the current anchors and nearby Mathlib lemmas. Keep the substitutions that Lean infers, the local hypotheses, and any open subgoals that a partial proof leaves.
\item \textbf{Extend the graph.} Let the constructor grow the relevant part of the skeleton from available lemmas and anchors. It uses the open subgoals to choose which consequences to explore within a fixed budget. Add a candidate lemma only after Lean's kernel checks its proof term.
\item \textbf{Resume and reuse.} Resume search with the new lemmas and let Lean check the complete proof. Keep verified lemmas available for later goals and track which ones those proofs reuse.
\end{enumerate}

Freeze the initial anchors before revealing the test goals. Run the search with Mathlib alone, with Mathlib and the frozen anchors, and with Mathlib and local growth from those anchors. Give all three the same LLM guidance, search procedure, and total budget, including model time. Count a goal when Lean checks its complete proof. Trace the anchors in that proof, disable each in turn, and record whether the same search still closes the goal and how long it takes. To test reuse, keep a new anchor and try it on a goal from another family.

\section*{10 Reproducibility}\label{reproducibility}
\addcontentsline{toc}{section}{10 Reproducibility}

Appendix I lists the primary schemata and their ordered inference pairs. From the root of the repository linked below, the following command reconstructs all 16 banks from their growth traces and checks the external proof scores.
\begin{quote}\small\ttfamily python3 release/propositional/verify.py -{}-reselect\end{quote}
The data record the source label, proof-graph digest, and anchors used in each shortened proof. Appendix G documents the separate blind comparison on 963 Metamath targets.

\section*{Statements and Declarations}

\subsection*{Data availability}\label{data-availability}
\addcontentsline{toc}{subsection}{Data availability}

The propositional release at \href{https://github.com/nnevdokimov/certified-proof-skeleton/tree/55fade4cb51b031403b99b3069650fbbce5b1b50/release/propositional}{\nolinkurl{github.com/nnevdokimov/certified-proof-skeleton}} (revision \texttt{55fade4cb}) supplies the constructor inputs, native and SGCD growth traces, certified banks, external proof graphs, and replay command. The arXiv source package also contains the same proof-level data in \texttt{anc/skeleton-data.json}. The \href{https://github.com/nnevdokimov/certified-proof-skeleton/tree/c6767d7eeaa7be42a39630c5a46e70873a8c01ae/release/article1}{earlier native release} documents the locked Metamath holdout and network profiles. The external inputs come from the Aviary package \citep{TetleyAviary} and \href{http://cs.christophwernhard.com/cdtools/exp-thgen/}{Wernhard's proof archive}. Third-party inputs retain their original licence terms.

\subsection*{Conflicts of interest}\label{conflicts-of-interest}
\addcontentsline{toc}{subsection}{Conflicts of interest}

The authors declare no conflict of interest concerning this article.

\subsection*{Funding}\label{funding}
\addcontentsline{toc}{subsection}{Funding}

The HSE University Basic Research Program funded this research {[}HSE-BR-2025-025{]}.

\subsection*{Acknowledgments}\label{acknowledgments}
\addcontentsline{toc}{subsection}{Acknowledgments}

We used the HSE University supercomputer to conduct the computational research.

\setlength{\bibsep}{0.7em}

\end{document}


\title[Direct cognition problem: a semantic skeleton of mathematics and its growth]{Direct cognition problem: a semantic skeleton of mathematics and its growth}
\subtitle{Supporting appendices}
\author[1]{\fnm{Vasilii A.} \sur{Gromov}}\email{stroller@rambler.ru}
\author*[1]{\fnm{Nikolai N.} \sur{Evdokimov}}\email{nevdokimov@hse.ru}
\affil[1]{\orgdiv{Laboratory for Semantic Analysis}, \orgname{National Research University Higher School of Economics}, \orgaddress{\street{11 Pokrovsky Bulvar}, \city{Moscow}, \postcode{109028}, \country{Russia}}}
\maketitle
\noindent\small ORCID (Nikolai N. Evdokimov): \url{https://orcid.org/0009-0003-3756-8281}\par\normalsize

\section*{Appendix A. Combinatorial bounds and resources}\label{appendix-a.-combinatorial-bounds-and-resources}
\addcontentsline{toc}{section}{Appendix A. Combinatorial bounds and resources}

\subsection*{A.1. Resources and notation}\label{a.1.-resources-and-notation}
\addcontentsline{toc}{subsection}{A.1. Resources and notation}

Resource accounting distinguishes the active representation \(M_{S}\), reserve \(M_{P}\), archive \(M_{\Omega}\), and peak process memory. Moving a proof to disk does not discard knowledge. Equal numbers of anchors therefore imply neither equal certificate sizes nor equal memory use (Table S1).

\begingroup
\small
\setlength{\tabcolsep}{3pt}
\renewcommand{\arraystretch}{1.13}

\begin{longtable}[]{@{}
  >{\raggedright\arraybackslash}p{(\linewidth - 2\tabcolsep) * \real{0.1707}}
  >{\raggedright\arraybackslash}p{(\linewidth - 2\tabcolsep) * \real{0.8293}}@{}}
\caption{Notation, mathematical objects, and resource units}\label{tab:8}\tabularnewline
\toprule\noalign{}
\begin{minipage}[b]{\linewidth}\raggedright
\textbf{Notation}
\end{minipage} & \begin{minipage}[b]{\linewidth}\raggedright
\textbf{Object and unit}
\end{minipage} \\
\midrule\noalign{}
\endfirsthead
\toprule\noalign{}
\begin{minipage}[b]{\linewidth}\raggedright
\textbf{Notation}
\end{minipage} & \begin{minipage}[b]{\linewidth}\raggedright
\textbf{Object and unit}
\end{minipage} \\
\midrule\noalign{}
\endhead
\bottomrule\noalign{}
\endlastfoot
\(|F|_{AST}\) & Number of all syntax-tree vertices; metric for the main bounds \(L = 25,200\) \\
\(\ell_{leaf}(F)\) & Number of leaves; metric for the separate implicational KSP benchmark \\
\(H_{d}^{sch}\) & Canonical schemata through depth \(d\) under the declared constraints \\
\(H_{4}\) & Small exact schematic benchmark through depth 4, with every stored schema of size at most 25 \\
H14 & Name of an earlier configuration, not depth 14 or benchmark \(H_{4}\) \\
\(\Omega_{0},\Omega_{b}\) & Common initial archive and archive of a verified continuation \\
\end{longtable}

\endgroup

The relation between the two size metrics depends on the language. A purely implicational formula satisfies \(|F|_{AST} = 2\ell_{leaf}(F) - 1\), whereas negation or modality invalidates this equality. Budget \(b\) covers the declared computational scope; accounting records the historical cost of constructing \(\Omega_{0}\) separately.

\subsection*{A.2. Combinatorial costs and finite approximation}\label{a.2.-combinatorial-costs-and-finite-approximation}
\addcontentsline{toc}{subsection}{A.2. Combinatorial costs and finite approximation}

\subsubsection*{A.2.1. Number of potential pairs}\label{a.2.1.-number-of-potential-pairs}
\addcontentsline{toc}{subsubsection}{A.2.1. Number of potential pairs}

Two-premise modus ponens admits an exact count of potential pairs before filtering. Let \(V_{\leq d}\) contain the available formulae, \(F_{d}\) denote the new frontier, \(I_{\leq d}\) denote the implicational major premises, and \(F_{d}^{I} = F_{d} \cap I_{\leq d}\). Restricting attention to pairs with at least one new premise gives

\[\fitdisplay{P_{d} = |F_{d}|\,|I_{\leq d}| + |V_{< d}|\,|F_{d}^{I}|.}\]

The two terms divide the ordered pairs by premise role. The first term counts a new minor premise, including pairs with two new premises; the second counts a new major premise and an old minor premise. This equality describes Cartesian enumeration under that division of roles; structural indexing reduces actual checks, whereas other rules require their own counters.

The potential-pair count satisfies \(P_{d} \leq N_{d}^{2}\) for \(N_{d} = |V_{\leq d}|\). A single check also incurs costs for formula size, unification, and side conditions. The inequality thus bounds pairs, not the generator's total running time, and schemata with several matching results can produce more conclusions than their pair count alone suggests.

\subsubsection*{A.2.2. Exponential size of the explicit representation}\label{a.2.2.-exponential-size-of-the-explicit-representation}
\addcontentsline{toc}{subsubsection}{A.2.2. Exponential size of the explicit representation}

An explicit-space lower bound requires a precise vertex convention. In the following result, each vertex represents a syntactically distinct formula, not a logical-equivalence class or all instances of one schema. We use a language with one fixed propositional letter \(p\), implication, and negation, and count abstract-syntax-tree vertices in \(|F|\); the calculus contains modus ponens and the schemata

\[\fitdisplay{\begin{matrix}
K(A,B) = A \rightarrow (B \rightarrow A), \\
S(A,B,C) = \lbrack A \rightarrow (B \rightarrow C)\rbrack \rightarrow \lbrack(A \rightarrow B) \rightarrow (A \rightarrow C)\rbrack
\end{matrix}}\]

The lower bound needs no additional axioms, including neither the third classical schema nor modal axioms. Let \(T_{L}\) denote the formulae admitting a proof in which \textbf{every} formula, including each axiom instance, contains at most \(L\) syntax-tree vertices. This definition expands axiom-schema substitutions explicitly and keeps the letter \(p\) fixed in the resulting formulae.

\textbf{Theorem 1 (full statement): size of the explicit proof space.} For every integer \(m \geq 0\),

\[\fitdisplay{|T_{18m + 17}|\  \geq \ {Cat}_{m},\quad\quad{Cat}_{m} = \frac{1}{m + 1}\binom{2m}{m}.}\]

These \({Cat}_{m}\) conclusions all contain \(4m + 3\) syntax-tree vertices. Each admits a proof with at most two modus-ponens applications along any path, counting axiom instances at height zero. As \(m \rightarrow \infty\), their number satisfies

\[\fitdisplay{{Cat}_{m} \sim \frac{4^{m}}{\sqrt{\pi}\, m^{3/2}}.}\]

\textbf{Proof.} Count fully parenthesised implicational formulae with \(m\) occurrences of \(\rightarrow\) and \(m + 1\) leaves that each carry the label \(p\). Let \(\mathcal{B}_{m}\) contain all such formulae. Splitting at the left and right subtrees gives

\[\fitdisplay{|\mathcal{B}_{0}| = 1,\quad\quad|\mathcal{B}_{m}| = \sum_{j = 0}^{m - 1}|\mathcal{B}_{j}|\,|\mathcal{B}_{m - 1 - j}|.}\]

This recurrence gives the standard Catalan enumeration of ordered full binary trees \citep{FlajoletSedgewick2009}. For each \(F \in \mathcal{B}_{m}\), set \(I_{F} = F \rightarrow F\); the map \(F \mapsto I_{F}\) injects the family into distinct conclusions, and the common leaf label prevents variable renaming from merging different tree shapes. With \(s = |F| = 2m + 1\), the size of each conclusion equals \(|I_{F}| = 2s + 1 = 4m + 3\).

The following derivation proves \(I_{F}\) and exposes every intermediate formula. Its two modus-ponens steps use only instances of \(K\) and \(S\). We can therefore check the size bound directly:

\[\fitdisplay{\begin{matrix}
\alpha = F \rightarrow \lbrack(F \rightarrow F) \rightarrow F\rbrack,\quad\quad K(F,F \rightarrow F), \\
\beta = \alpha \rightarrow \{\lbrack F \rightarrow (F \rightarrow F)\rbrack \rightarrow (F \rightarrow F)\},\quad\quad S(F,F \rightarrow F,F), \\
\gamma = \lbrack F \rightarrow (F \rightarrow F)\rbrack \rightarrow (F \rightarrow F),\quad\quad MP(\alpha,\beta), \\
\delta = F \rightarrow (F \rightarrow F),\quad\quad K(F,F), \\
I_{F} = F \rightarrow F,\quad\quad MP(\delta,\gamma).
\end{matrix}}\]

The respective sizes of the five formulae equal \(4s + 3\), \(9s + 8\), \(5s + 4\), \(3s + 2\), and \(2s + 1\). Their maximum equals \(9s + 8 = 18m + 17\), so every \(I_{F}\) belongs to \(T_{18m + 17}\). Applying Stirling's formula to \({Cat}_{m}\) gives the asymptotic estimate. \(\square\)

The theorem bounds output size, without claiming a new complexity result for finding an individual proof. The standard identity derivation and Catalan count provide an explicit witness; the displayed derivation supplies the constant \(18\), without any claim of optimality. For \(L \geq 17\), setting \(m = \lfloor(L - 17)/18\rfloor\) yields

\[\fitdisplay{|T_{L}| \geq {Cat}_{\lfloor(L - 17)/18\rfloor} = \Omega\left( \frac{2^{L/9}}{L^{3/2}} \right)\quad(L \rightarrow \infty).}\]

\subsubsection*{A.2.3. Limits of complete storage under a fixed resource bound}\label{a.2.3.-limits-of-complete-storage-under-a-fixed-resource-bound}
\addcontentsline{toc}{subsubsection}{A.2.3. Limits of complete storage under a fixed resource bound}

\textbf{Corollary 1.} An algorithm that materialises and retains a separate record for every formula in \(T_{L}\) requires \(\Omega({Cat}_{m})\) records and output operations when \(L = 18m + 17\). If each record contains a separate syntax tree without sharing, the family's conclusions alone require \(\Omega(m{Cat}_{m})\) symbols. Sharing subexpressions can reduce the second bound, but cannot reduce the number of explicitly addressable vertices under this model.

A store with capacity for at most \(M\) such records cannot retain a complete materialised graph for unboundedly increasing \(L\). Choosing \(m\) with \({Cat}_{m} > M\) exceeds that capacity, which covers \textbf{the entire retained explicit representation}, not merely RAM; unlimited disk does not count as a fixed resource. Likewise, with at most \(r\) new records per epoch and completeness at epoch \(t\) for \(L_{t} = 18t + 17\), the necessary inequality \({Cat}_{t} \leq M_{0} + rt\) fails for sufficiently large \(t\).

\textbf{Proof.} The theorem's pairwise distinct conclusions each require a separate record and at least one output event for that record. The Catalan numbers grow faster than every linear function of \(t\). These requirements establish both the storage obstruction and the epoch-rate obstruction. \(\square\)

The corollary limits complete retained output, not every form of enumeration. It permits infinite enumeration with growing time, streamed output without retaining every result, and ever slower increases in the size bound. Nor does it imply that a finite bounded closure lacks a computation: a fixed finite alphabet and size bound give a finite instance space, so exhaustive saturation with decidable rules and deduplication terminates.

\subsubsection*{A.2.4. Limits of transferring the bound to a schematic generator}\label{a.2.4.-limits-of-transferring-the-bound-to-a-schematic-generator}
\addcontentsline{toc}{subsubsection}{A.2.4. Limits of transferring the bound to a schematic generator}

Theorem 1 does \textbf{not} establish exponential vertex growth for our schematic generator. Every \(I_{F}\) in its proof instantiates the single schema \(A \rightarrow A\), whereas the generator stores canonical schemata and most-general unification results. The finite schematic benchmark \(H_{d}\) and the explicit-instance set \(T_{L}\) therefore describe different objects, just as proof height, formula size, and epoch number measure different quantities.

Schematic modus ponens yields an unbounded exact graph under the conditions below. The construction below supplies distinct schemata at successive depths. Its depth-dependent lower bound grows only linearly.

\textbf{Theorem 2 (restatement): unboundedness of the exact schematic graph.} Let \(H_{d}^{sch}\) contain the initial schema \(K\) and every most-general result of two-premise modus ponens through depth \(d\), with separate variable renaming for the premises and canonicalisation solely up to renaming. Impose no size bound, subsumption, width limit, or pruning of rule applications. Then

\[\fitdisplay{|V(H_{d}^{sch})| \geq d + 1.}\]

Consequently, a fixed capacity of \(M\) vertices cannot retain the complete exact schematic graph indefinitely. Under a size bound, the same estimate holds through depth \(d\) whenever the bound for \textbf{stored canonical schemata} admits every initial schema and reaches at least \(2d + 5\). This condition does not bound the explicitly expanded substitution instances of the major premise; to impose that further requirement, we would need to calculate those sizes separately.

\textbf{Proof.} Construct a family by adding a fresh antecedent at each step. Set \(Q_{0} = A \rightarrow (B \rightarrow A)\) and choose \(X_{n + 1}\) outside the variables of \(Q_{n}\). Define

\[\fitdisplay{Q_{n + 1} = X_{n + 1} \rightarrow Q_{n}.}\]

Use \(Q_{n}\) as the minor premise and a separate copy of \(K(U,V) = U \rightarrow (V \rightarrow U)\) as the major premise. After separate renaming, their most-general unifier assigns \(U: = Q_{n}\) and leaves \(V\) free, yielding \(V \rightarrow Q_{n}\text{,}\) which agrees with \(Q_{n + 1}\) after canonicalisation. Induction therefore gives \(Q_{n} \in H_{n}^{sch}\); their distinct, renaming-invariant sizes \(|Q_{n}| = 2n + 5\) distinguish the schemata, so \(Q_{0},\ldots,Q_{d}\) alone require \(d + 1\) vertices. \(\square\)

Theorem 2 concerns exact exhaustive search and does not promise that a particular bounded policy will obtain the whole family. A fixed size bound truncates the family, whereas subsumption changes the identification relation. Exponential growth of \(|H_{d}^{sch}|\) thus remains a separate question, which neither observed jumps in finite sizes nor the quadratic potential-pair counter resolves.

The lower bound motivates avoiding explicit enumeration of every instance, not deleting arbitrary schemata. A compact skeleton needs an additional empirical criterion: does it retain opportunities to derive further verifiable results? Agreement with human lemmas helps assess that choice but does not follow logically from the theorem.

\subsubsection*{A.2.5. Cost accounting}\label{a.2.5.-cost-accounting}
\addcontentsline{toc}{subsubsection}{A.2.5. Cost accounting}

We measure how long the generator spends on each operation: indexing, matching, canonicalisation, checking, selection, and storage. We record the total as

\[\fitdisplay{T_{d} = T_{d}^{index} + T_{d}^{match} + T_{d}^{canon} + T_{d}^{check} + T_{d}^{select} + T_{d}^{store}.}\]

Separate counters track potential pairs, actual attempts, successful applications, new formulae, and new proof occurrences. The share of new formulae among the results can change over time. More rediscoveries can therefore increase computational cost without adding comparable knowledge.

Activating fewer premises limits part of the enumeration, without establishing linear complexity or a constant total active-set size. The passive reserve, archive, unification candidates, and major premises can continue to grow. A width-only estimate in \(W\) therefore falls short: experiments must measure whole-process memory and the cost of generating rejected candidates.

\subsubsection*{A.2.6. What the skeleton approximates}\label{a.2.6.-what-the-skeleton-approximates}
\addcontentsline{toc}{subsubsection}{A.2.6. What the skeleton approximates}

We do not call a finite skeleton a numerical approximation to the `full semantics of mathematics' without defining an error. Instead, the small finite benchmark measures losses of accessible formulae and their recoverability, whereas direct continuation measures new formulae relative to the common archive, external human coverage, and fixed-generator costs. Retaining almost every easy formula does not guarantee that search retains rare useful lemmas.

For an exact closure \(C\) and initial set \(\Gamma\), we measure retained non-axiomatic growth through direct coverage. The numerator counts discovered non-axiomatic vertices within the exact closure, and the denominator counts that closure's non-axiomatic vertices. Define

\[\fitdisplay{R_{D} = \frac{|(V_{found}\setminus \Gamma) \cap (C\setminus \Gamma)|}{|C\setminus \Gamma|}.}\]

A zero denominator leaves this measure undefined. A separate measurement records recovered coverage after a fixed extra budget. Estimates from a few generations do not establish uncomputability or unattainability of a fixed finite slice.

\section*{Appendix B. Guarantees for the proof representation}\label{appendix-b.-guarantees-for-the-proof-representation}
\addcontentsline{toc}{section}{Appendix B. Guarantees for the proof representation}

\subsection*{B.1. Soundness of transfer between epochs}\label{b.1.-soundness-of-transfer-between-epochs}
\addcontentsline{toc}{subsection}{B.1. Soundness of transfer between epochs}

\textbf{Proposition 1.} Assume sound initial rules, a checkable proof from the initial base for every transferred lemma, and well-founded certificate dependencies. Also assume that substituting those proofs for lemma references preserves contexts, types, and every side condition. Every result of finitely many epochs through sound applications of these rules and lemmas then admits a proof from the initial base.

\textbf{Proof.} Order the certificates topologically by dependency. The first lemma expands using only initial rules; if all predecessors admit finite sound expansions, substituting those expansions for their references gives a finite sound proof of the next lemma. Induction covers every epoch result, with each application checking its side conditions and no transferred lemma acquiring a stronger interface than its certificate permits. \(\square\)

The argument requires well-founded dependencies. Cyclic references without a finite initial witness do not constitute a proof. The proposition therefore establishes soundness, without any claim about the efficiency of lemma selection.

\subsection*{B.2. Preservation of existing proofs}\label{b.2.-preservation-of-existing-proofs}
\addcontentsline{toc}{subsection}{B.2. Preservation of existing proofs}

\textbf{Proposition 2.} Deactivation preserves every previously checked proof if it leaves the stored certificates and their required dependencies unchanged. Activity status contributes no logical condition to checking an archived inference. This guarantee concerns existing proofs, not future search results.

\textbf{Proof.} Checking an archived inference depends on its certificate and necessary dependencies, not on whether search currently activates its formula. Deactivation under the stated hypothesis changes none of those inputs. Every previously checked proof therefore retains its validity. \(\square\)

The proposition does not preserve the set of future results. A system in which a deactivated lemma later supplies the only available premise for a new rule application already refutes that stronger claim. A sound archive therefore does not guarantee that future search retains those opportunities.

\subsection*{B.3. Lemma expansion and its resource cost}\label{b.3.-lemma-expansion-and-its-resource-cost}
\addcontentsline{toc}{subsection}{B.3. Lemma expansion and its resource cost}

The size of the largest intermediate formula provides one resource measure for a proof. Let \(\ell(\varphi)\) denote formula size, and minimise that maximum over proofs of a provable statement \(v\) in calculus \(\mathcal{C}\). This gives

\[\fitdisplay{b_{\mathcal{C}}(v) = \min_{\pi \vdash_{\mathcal{C}}v}\max_{\psi \in \pi}\ell(\psi).}\]

Adding a ready-made provable lemma preserves unrestricted provability but can lower the required resource bound and proof height. For a set \(\Lambda\) of added provable lemmas, \(b_{\mathcal{C} + \Lambda}(v) \leq b_{\mathcal{C}}(v)\). A skeleton can therefore offer computational value without strengthening the original logic \citep{CookReckhow1979}.

\textbf{Proposition 3.} Suppose every used instance of an added lemma admits a proof in the base calculus with maximum formula size at most \(\kappa\) and height at most \(h\). Consider a proof of height \(d\), with intermediate formula sizes at most \(m\), that treats these lemmas as initial premises. Expanding those lemmas gives a base-calculus proof with size bound \(max\{ m,\kappa\}\) and height at most \(d + h\).

\textbf{Proof.} Replace each leaf use of an added lemma by its proof. The size of each new formula does not exceed \(\kappa\), all other formulae remain unchanged, and each path from root to original leaf gains at most \(h\) steps. These bounds concern certificates after full expansion into the common base system; nested references across epochs first require those full bounds, rather than repeated resets of local depth. \(\square\)

The purely implicational KSP calculus, with schemata \(K\), \(S\), and Peirce, admits a sharper comparison in a different metric. Use the \textbf{leaf} metric \(\ell_{leaf}\) and its associated quantity \(b_{leaf}\). Adding the ready-made schema \(I:A \rightarrow A\) gives

\[\fitdisplay{b_{leaf}^{KSP + I}(v) \leq b_{leaf}^{KSP}(v) \leq \frac{9}{2}\, b_{leaf}^{KSP + I}(v).}\]

Eliminating \(I\) increases height by at most two. The formula \(p \rightarrow p\) attains the factor \(9/2\): its conclusion contains two leaves, whereas every base proof minimising the largest leaf count reaches nine leaves; counting all tree vertices instead gives 3 and 17, so that metric does not share the factor. The lower bound relies on a complete computational closure check through eight leaves, and Appendix C, Section C.1 presents the construction; this result bounds resources in one calculus, not search-time speed-up.

\subsection*{B.4. Completeness and bounded resources}\label{b.4.-completeness-and-bounded-resources}
\addcontentsline{toc}{subsection}{B.4. Completeness and bounded resources}

\textbf{Proposition 4.} Fix a finite complete registry of admissible rule applications and determine availability from every accumulated proven formula, not merely the current active set. Assume that the process eventually executes every application once all its premises occur. The process then reaches the least deductive closure.

\textbf{Proof.} Induct on finite proof height. Once all premises occur, the required application becomes available, and fairness eventually schedules it; the finite registry consequently reaches all derivable vertices in finitely many steps. Bounded width and irreversible pruning can violate that fairness condition, so resource-bounded algorithms do not inherit completeness automatically. \(\square\)

\subsection*{B.5. Why the greedy library never increases the training cost}\label{b.5.-why-the-greedy-library-never-increases-the-training-cost}
\addcontentsline{toc}{subsection}{B.5. Why the greedy library never increases the training cost}

\textbf{Lemma 1 (library expansion and non-negative training gain).} For a finite acyclic archive, the library encoding expands into the original forest of first proofs. If the selector starts from the empty library and admits only strictly positive increments of \(G\), the final gain on the training forest remains non-negative. Neither claim extends that gain to future proofs.

\textbf{Proof.} The parents of every reference vertex precede it in topological order. The codec therefore finishes expanding the definitions and recovers the original rule applications with the same premise roles. The checker verifies the original certificates and checks that every expansion remains admissible, establishing soundness. For the empty library, \(G = 0\), and the sum of positive increments remains non-negative; no step of this argument uses future proofs. \(\square\)

The lemma guarantees neither gain in later epochs nor faster search. The metric counts symbols in the expanded proof term, not bytes or time, whereas the common archive already shares identical subproofs. Consequently, \(G\) measures no additional compression of that directed acyclic graph.

\section*{Appendix C. Additional checks}\label{appendix-c.-additional-checks}
\addcontentsline{toc}{section}{Appendix C. Additional checks}

\subsection*{C.1. Resource cost of a ready-made identity}\label{c.1.-resource-cost-of-a-ready-made-identity}
\addcontentsline{toc}{subsection}{C.1. Resource cost of a ready-made identity}

The identity calculation uses \(J = X \rightarrow X\), \(P = X \rightarrow (J \rightarrow X)\), and \(Q = X \rightarrow (X \rightarrow X)\). The following proof derives the identity through two applications of modus ponens. Its explicit intermediate formulae provide the resource accounting below:

\[\fitdisplay{\begin{matrix}
1. & P & K, \\
2. & P \rightarrow (Q \rightarrow J) & S, \\
3. & Q \rightarrow J & MP(1,2), \\
4. & Q & K, \\
5. & J & MP(4,3).
\end{matrix}}\]

This calculation uses only the leaf metric \(\ell_{leaf}\). If \(X\) contains \(n\) variable occurrences, the sizes of the five formulae equal \(4n\), \(9n\), \(5n\), \(3n\), and \(2n\), and the proof reaches height two when axioms start at height zero. Using \(X \rightarrow X\) within a window of size \(m\) requires \(n \leq \lfloor m/2\rfloor\), so expansion fits within

\[\fitdisplay{\kappa(m) = max\{ m,9\lfloor m/2\rfloor\}.}\]

The complete closure through eight leaves passes checking but contains no \(p \rightarrow p\), whereas the proof above fits within nine leaves. The lower bound therefore follows from a finite computational check of this particular calculus, not merely from the upper-bound construction; the linked report retains checksums and independent checking of the smaller window. The factor \(9/2\) does not automatically transfer to other axiomatisations, search time, or certificate length.

\section*{Appendix D. Continuation and Pareto evidence}\label{appendix-d.-continuation-and-pareto-evidence}
\addcontentsline{toc}{section}{Appendix D. Continuation and Pareto evidence}

\subsection*{D.1. Frozen pair selection and replay panel}\label{d.1.-frozen-pair-selection-and-replay-panel}
\addcontentsline{toc}{subsection}{D.1. Frozen pair selection and replay panel}

The panel crosses six policies, three logic--size regimes, and scheduler states 8701--8708. The policies admit all compatible pairs, a random half, or a learned half; the latter two also run with deferred replay, and a sixth variant retains a learned quarter with replay. All conditions use the same initial axioms, canonical forms, model files, lifecycle, and certificate checker within a regime. The original training archives supply labels and weights; this panel introduces no human theorem targets or new model fitting.

Every source receives nominal readings at 60, 180, and 600 CPU-seconds. Each reading supplies anchor caps 4, 16, and 64, and each resulting package receives a separate neutral continuation with readings at 10, 30, and 60 CPU-seconds. This design produces 144 sources, 1,296 continuations, and 3,888 continuation readings. Nested readings and repeated source packages do not count as independent trajectories. The prespecified primary endpoint uses source ceiling 180, cap 16, and continuation 30; the main text retains that endpoint rather than selecting a more favourable secondary cell.

The model describes only already available premises. A fixed share of exploratory candidates accompanies learned ranking, and admitted pairs preserve the generator's traversal order. Replay stores up to 32,768 skipped pairs, attempts up to 128 ready pairs per epoch, and reserves at most four ordinary premise slots. Queue overflow counts rejected offers rather than distinct lost pairs or theorems. None of these finite quotas guarantees fair exhaustive enumeration.

The independent cluster checker verifies 24 logic--state blocks. Local checks compare all 229,221 journal-chain files against those checked records and replay every one of the 1,296 exported proof-backed packages. Those package replays traverse 172,522 support records with repetitions across packages. Transport checks link the downloaded journals to the cluster audit; they do not replay every elementary proof locally.

The reported source, selection, and continuation components total 27.17 CPU-hours; including the two audit stages gives 35.02 CPU-hours, excluding some startup and orchestration. Earlier training and preparation cost 37.79 CPU-seconds once. These components describe measured process work rather than the SLURM bill or a hardware-independent complexity measure. At the credited endpoints, memory guards and incomplete layers can leave actual CPU below the assigned ceiling.

\subsection*{D.2. Finite grid and Pareto coordinates}\label{d.2.-finite-grid-and-pareto-coordinates}
\addcontentsline{toc}{subsection}{D.2. Finite grid and Pareto coordinates}

The new descriptive Pareto check uses only source ceiling 180 and continuation horizon 30. Each regime contains six policies at three anchor caps, hence 18 mean triples across eight schedules. The coordinates use credited chain CPU, complete package bytes, and new schemata outside the concurrent source union and old training. We recompute dominance from the existing aggregate records; no search, hypothesis test, or selection of favourable states follows this check.

The frontiers contain five settings at propositional size 25, seven at propositional size 200, and five at modal size 25. The table lists their mean coordinates. Names H and Q denote learned half and quarter selection, R denotes random half selection, and +replay denotes deferred attempts. The absence of a policy from this finite mean frontier does not imply dominance at every schedule, every cost, or another horizon (Table S2).

\begingroup
\small
\setlength{\tabcolsep}{3pt}
\renewcommand{\arraystretch}{1.13}

\begin{longtable}[]{@{}
  >{\raggedright\arraybackslash}p{(\linewidth - 10\tabcolsep) * \real{0.1650}}
  >{\raggedright\arraybackslash}p{(\linewidth - 10\tabcolsep) * \real{0.1651}}
  >{\raggedleft\arraybackslash}p{(\linewidth - 10\tabcolsep) * \real{0.1651}}
  >{\raggedleft\arraybackslash}p{(\linewidth - 10\tabcolsep) * \real{0.1651}}
  >{\raggedleft\arraybackslash}p{(\linewidth - 10\tabcolsep) * \real{0.1744}}
  >{\raggedleft\arraybackslash}p{(\linewidth - 10\tabcolsep) * \real{0.1651}}@{}}
\caption{Mean CPU, package bytes, and new schemata for the finite three-coordinate Pareto frontiers}\label{tab:31}\tabularnewline
\toprule\noalign{}
\begin{minipage}[b]{\linewidth}\raggedright
\textbf{Regime}
\end{minipage} & \begin{minipage}[b]{\linewidth}\raggedright
\textbf{Policy}
\end{minipage} & \begin{minipage}[b]{\linewidth}\raggedleft
\textbf{Anchor cap}
\end{minipage} & \begin{minipage}[b]{\linewidth}\raggedleft
\textbf{Chain CPU s}
\end{minipage} & \begin{minipage}[b]{\linewidth}\raggedleft
\textbf{Package bytes}
\end{minipage} & \begin{minipage}[b]{\linewidth}\raggedleft
\textbf{New schemata}
\end{minipage} \\
\midrule\noalign{}
\endfirsthead
\toprule\noalign{}
\begin{minipage}[b]{\linewidth}\raggedright
\textbf{Regime}
\end{minipage} & \begin{minipage}[b]{\linewidth}\raggedright
\textbf{Policy}
\end{minipage} & \begin{minipage}[b]{\linewidth}\raggedleft
\textbf{Anchor cap}
\end{minipage} & \begin{minipage}[b]{\linewidth}\raggedleft
\textbf{Chain CPU s}
\end{minipage} & \begin{minipage}[b]{\linewidth}\raggedleft
\textbf{Package bytes}
\end{minipage} & \begin{minipage}[b]{\linewidth}\raggedleft
\textbf{New schemata}
\end{minipage} \\
\midrule\noalign{}
\endhead
\bottomrule\noalign{}
\endlastfoot
Prop. 25 & H & 4 & 81.02 & 2347 & 21818.50 \\
Prop. 25 & H+replay & 4 & 169.22 & 2041 & 23390.25 \\
Prop. 25 & H & 16 & 81.62 & 3195 & 22820.12 \\
Prop. 25 & H+replay & 16 & 169.53 & 3145 & 27285.12 \\
Prop. 25 & H & 64 & 82.36 & 3385 & 25771.12 \\
Prop. 200 & R & 4 & 140.12 & 2522 & 22841.25 \\
Prop. 200 & R+replay & 4 & 144.59 & 2202 & 21137.38 \\
Prop. 200 & H+replay & 4 & 136.12 & 2604 & 26015.50 \\
Prop. 200 & Q+replay & 4 & 142.46 & 2957 & 28446.88 \\
Prop. 200 & H+replay & 16 & 136.06 & 3487 & 26805.12 \\
Prop. 200 & Q+replay & 16 & 142.67 & 4115 & 28640.75 \\
Prop. 200 & H+replay & 64 & 137.39 & 3932 & 27580.38 \\
Modal 25 & H & 4 & 41.37 & 1283 & 21918.12 \\
Modal 25 & H+replay & 4 & 140.64 & 2236 & 29698.62 \\
Modal 25 & H+replay & 16 & 139.00 & 2825 & 25456.50 \\
Modal 25 & R+replay & 64 & 175.00 & 3848 & 30449.50 \\
Modal 25 & H+replay & 64 & 141.78 & 3231 & 30158.75 \\
\end{longtable}

\endgroup

These rows expose a trade-off. A learned half-filter without replay reaches a cheaper region at size 25, whereas replay reaches greater novelty or smaller proof-backed packages in other cells. At propositional size 200, half and quarter replay both retain nondominated settings. We therefore retain the finite empirical Pareto set rather than nominate one universal skeleton or collapse all resource coordinates into an unexplained scalar score.

\section*{Appendix E. Native proof networks}\label{appendix-e.-native-proof-networks}
\addcontentsline{toc}{section}{Appendix E. Native proof networks}

\subsection*{E.1. Native basis and search}\label{e.1.-native-basis-and-search}
\addcontentsline{toc}{subsection}{E.1. Native basis and search}

The native implementation reads the same pinned \path{set.mm} source as the preceding study, at commit \path{584b6852f57f281e9be1793332f1e52e077693c8}, with SHA-256 \path{90de6bac87d549023fc83fb698a1baacfed3f99b71438722171674f16777c881}. The basis comprises 25 logical or ZFC assertions (including modus ponens and generalisation), six definitions, and 12 syntax assertions. Variables carry the \path{wff}, \path{setvar}, or \path{class} type. Canonical statement identity includes the formula and its mandatory distinct-variable frame; different such frames can therefore create separate vertices for one formula body. Certificates retain substitutions, joint premises, and the additional variables that proof replay needs.

The native given-clause policy selects one waiting statement and combines it with every previously processed statement, including admissible self-detachment. Modus ponens uses typed unification and preserves all required distinct-variable conditions. Generalisation considers each set variable already present in the premise plus one fresh representative. The age/weight schedule takes the oldest waiting statement every fifth non-source choice and otherwise cycles through propositional, predicate-logical, and set-theoretic families. We assign these families according to syntax and source origins, without ranking mathematical importance. A contraction priority favours a conclusion shorter than at least one parent. One-way typed subsumption checks its substitution witness and distinct-variable frame before treating a statement as search-redundant. These operations preserve archived proofs; finite bounds can still prevent fair enumeration.

The run allows 300 search CPU-seconds, 300,000 archived nodes, 10,000 selections, formula size at most 160, and proof height at most 64. It reaches the node guard after 222.310068 search CPU-seconds. Its archive contains 300,000 statement records and 259,344 distinct formula bodies. The processed core contains 2,679 derived anchors and 29 sources; all first-proof ancestors of those anchors already belong to that core. Its 2,708 records contain 2,452 distinct formula bodies and reach proof height 25. Official Metamath replay checks the exported compact library and its references to the original source. The policy computes many unselected candidates, so these results do not establish generation of skeleton points alone or an equal-budget advantage over another generator.

\subsection*{E.2. Finite reference and human fragment}\label{e.2.-finite-reference-and-human-fragment}
\addcontentsline{toc}{subsection}{E.2. Finite reference and human fragment}

The exact reference starts with 29 source schemata and exhausts the declared MGU-MP/generalisation operator through height three. Its successive cumulative sizes reach 29, 185, 1,273, and 11,772. It applies no active-core, formula-size, or skeleton-selection cut. Earlier generic CLI fields record size 160 and proof height 64, but the exact mode does not apply them; its largest observed formula reaches size 163. The separate specification records this distinction. Independent replay and a separate enumerator check the retained nodes and inference records.

The two reference projections share the same vertex order. The first records one accepted proof for each derived statement and contains 19,461 distinct directed premise arcs. The second records every enumerated inference and contains 262,655 directed arcs, or 242,726 unique undirected edges after collapsing direction and multiplicity. All first-proof arcs also occur in the second graph. Its cycles encode alternative derivations rather than circular certificates. The final layer contributes 10,499 vertices with no observed later use. These vertices lie on the finite frontier; their lack of later use therefore does not establish uselessness.

The human extractor follows every actual logical and syntactic proof dependency and excludes statements that require a source assertion outside the declared basis. It retains 2,308 logical assertions, including the two primitive rules; removing those rule vertices leaves the 2,306-vertex graph. It keeps open-context human assertions as contextual rules rather than unconditional native schemata. Independent Metamath replay validates the extracted proofs. The resulting fragment does not represent all of Metamath, and a common axiom basis does not equalise proof granularity, elapsed search, or depth.

The separate human-statement diagnostic merges duplicate closed statements into 919 canonical statements and treats this set separately from the preceding 108-statement propositional benchmark. It finds 34 one-way matches in the initial basis, 82 in the native skeleton, and 98 in its archive; exact matches reach 12, 55, and 66. Typed one-way matching keeps target variables rigid and checks required distinct-variable conditions. This post-hoc diagnostic supplies no unseen holdout and no causal efficiency comparison. The core contains implication identity and both double-negation directions, but still lacks equality reflexivity, equality symmetry, and curried implication transitivity. Its operational scope therefore falls short of a general human mathematical library.

\subsection*{E.3. Metrics and reference graphs}\label{e.3.-metrics-and-reference-graphs}
\addcontentsline{toc}{subsection}{E.3. Metrics and reference graphs}

The graph builder links every distinct premise assertion to each conclusion that cites it, removes loops and repeated arcs, and excludes primitive-rule vertices in all objects. For paths, clustering, and modularity, it collapses direction and parallel edges to a simple undirected graph. Local clustering contributes zero at degree below two. Breadth-first search from every vertex computes exact observed distances without retaining a dense distance matrix. Mean path and diameter refer to the largest component. Three fixed-seed Louvain starts supply the largest observed modularity for each graph.

Ten simple rewired controls preserve the undirected degree of every labelled vertex exactly. Each control attempts ten rewiring moves per edge; this finite procedure does not establish independent uniform sampling from the degree-constrained ensemble. The reference mean Q0 averages their Louvain modularities. We divide observed clustering by mean control clustering, then divide the result by the ratio of observed mean path to mean control path. Observed paths cover every source; control paths use 64 sampled sources. Under homogeneous G(n,m) controls, the first three graphs yield ratios 31.162, 17.091, and 294.915, whereas degree-preserving controls yield 3.990, 0.765, and 0.592. This contrast shows why a short path and an ER-based ratio alone do not isolate clustering beyond degree heterogeneity.

The directed out-degree counts distinct conclusions that cite a vertex. The pure-tail fit uses the exact discrete power-law likelihood and chooses its lower cutoff through the declared fit procedure. Each of 20,000 semiparametric bootstrap draws refits both the cutoff and exponent, with fixed seed 20261020 and no threshold-driven stopping. The test follows a conditional iid tail model, not independent proof-graph trajectories. 

The first-proof native graphs attain maximum core number two, the human named graph attains six, and the all-inference graph attains 190. These values do not rank mathematical depth. In a DAG that records only first proofs through binary modus ponens and unary generalisation, every derived vertex cites at most two earlier neighbours.

\textbf{Proof of the format bound.} Consider any nonempty induced subgraph and choose its last vertex in a topological order. It cannot point to another vertex in that subgraph, and at most two premise neighbours point to it. Every induced subgraph therefore contains a vertex of degree at most two, which bounds degeneracy and maximum core number by two. Named human steps can cite many premises, and the cyclic all-inference graph does not satisfy this first-proof DAG condition.

\section*{Appendix F. Native cumulative libraries and structural synthesis}\label{appendix-f.-native-cumulative-libraries-and-structural-synthesis}
\addcontentsline{toc}{section}{Appendix F. Native cumulative libraries and structural synthesis}

\subsection*{F.1. Exact gains above a nonempty dictionary}\label{f.1.-exact-gains-above-a-nonempty-dictionary}
\addcontentsline{toc}{subsection}{F.1. Exact gains above a nonempty dictionary}

Let \(L(D;R)\) denote the code length inside the square brackets in the definition of \(G(D;R)\). Fix a finite acyclic proof archive, a forest \(R\), and a mandatory base dictionary \(D_{0}\). References stop expansion at selected definitions; each selected definition expands its root once and charges a header. The codec recomputes expanded occurrences across both the forest and the current definition bodies.

\textbf{Proposition F.1.} Suppose an eligible unselected fragment \(v\) occurs \(r\) times in that current expansion and its current body contains \(b\) symbols. Adding \(v\) changes encoded length by

\[\fitdisplay{L(D;R) - L(D \cup \{ v\};R) = r(b - 1) - (b + 1) = (r - 1)(b - 1) - 2.}\]

If the codec adds only fragments with a strictly positive value and recalculates the expansion each time it adds a fragment, its final dictionary \(D_{*}\) satisfies

\[\fitdisplay{L(D_{*};R) \leq L(D_{0};R).}\]

\textbf{Proof.} Before the codec adds the fragment, the \(r\) occurrences each expand a body of length \(b\). Afterwards, each occurrence contributes one reference, whereas the new definition contributes one body and one header. The length change therefore equals \(r(b - 1) - (b + 1)\). Acyclic dependencies ensure finite expansion and preserve the represented forest. Summing the positive gains over all additions telescopes from \(L(D_{0};R)\) to \(L(D_{*};R)\). The finite archive supplies finitely many eligible roots, so this greedy procedure terminates. \(\square\)

This bound compares the final code with the existing code; it does not guarantee non-negative total gain against an empty dictionary. Mandatory entries may already carry overhead. The result also leaves physical DAG storage and future search cost outside its guarantee. The native implementation applies no target cardinality to the upper dictionary; a separate finite working-set limit governs source construction.

\subsection*{F.2. Checked deduction from open MP programmes}\label{f.2.-checked-deduction-from-open-mp-programmes}
\addcontentsline{toc}{subsection}{F.2. Checked deduction from open MP programmes}

An open programme contains ordered binary MP nodes and boundary hypotheses. A set partition identifies boundary hypotheses that must coincide. The compiler introduces fresh typed variables for the distinct holes and internal conclusions, imposes every MP equation jointly, and rejects an occurs-check or type failure. A successful unifier instantiates a finite consistent open derivation. Gen terminates the abstraction and never enters its open interior.

\textbf{Proposition F.2.} Let \(P\) denote a finite consistent derivation of \(C\) from boundary hypotheses \(H_{1},\ldots,H_{m}\), using MP and instances of ax-1 and ax-2. For every permutation \(\pi\), the original axioms derive

\[\fitdisplay{H_{\pi(1)} \rightarrow \left( H_{\pi(2)} \rightarrow \cdots \rightarrow (H_{\pi(m)} \rightarrow C)\cdots \right).}\]

\textbf{Proof.} Fix a hypothesis \(A\) and induct on \(P\). For a leaf that repeats \(A\), the standard ax-1/ax-2 derivation yields \(A \rightarrow A\). For an axiom or a hypothesis independent of \(A\), lift its conclusion \(B\) through ax-1 and MP to obtain \(A \rightarrow B\). For an MP step from \(B\) and \(B \rightarrow C\), the induction hypotheses give \(A \rightarrow B\) and \(A \rightarrow (B \rightarrow C)\). An instance of ax-2 and two MP applications yield \(A \rightarrow C\). Repeating this construction discharges the hypotheses in the reverse of the desired antecedent order and proves the displayed schema. All steps preserve the fixed substitutions; no step assumes unrestricted deduction across Gen.~\(\square\)

The compiler may reuse an existing more general closed schema through a checked one-way instance. It records explicit parent substitutions and mandatory distinct-variable obligations at every native step. Independent Metamath replay validates the exported result. This establishes correctness for the actual delivered outputs, not completeness of the bounded catalogue or a new deduction theorem.

\textbf{Corollary F.3.} Adding the compiled schemata conservatively extends the original axiom presentation. If the new interface also retains every earlier permitted statement and its applicability frame, its initial one-way coverage cannot decrease.

\textbf{Proof.} Proposition F.2 derives every added schema from the original axioms, so replacing its uses by those proofs eliminates every added lemma reference. Retaining each earlier permitted statement preserves every one of its admissible substitution instances. Both claims concern logical access, not the schedule or CPU that a search procedure spends exploring it. \(\square\)

A larger interface can dilute sampling and increase applicability checks. Hence Corollary F.3 does not order future coverage under a fixed budget. The common-residual and cold-cost endpoints test precisely that missing computational property.

\subsection*{F.3. Finite catalogue and source profile}\label{f.3.-finite-catalogue-and-source-profile}
\addcontentsline{toc}{subsection}{F.3. Finite catalogue and source profile}

Ordered binary trees with \(k\) internal MP nodes supply \({Cat}_{k}\) shapes and \(k + 1\) boundary positions. Enumerating their equality partitions and the discharge orders gives the elementary finite upper bound

\[\fitdisplay{\sum_{k = 0}^{q}{Cat}_{k}\,{Bell}_{k + 1}(k + 1)!}\]

on requested programmes before unification failures and duplicate formulae. This count does not bound the cost of all term substitutions or proof compilation. At \(q = 3\), the fixed builder delivers 217 distinct certified schemata under its stored-formula and certificate-height guards; 16 arise from the subgrammar through two MP nodes. Delivered schemata may cover several requested principal formulae.

For the adaptive score, the programme walker composes every parent substitution along an MP fragment. It compares the resulting boundary formulae in the root's common variable frame, then assigns the equality partition by first occurrence. It counts roots only in the newly built source segment and weights each occurrence by one plus the number of direct premise uses already present in that source DAG. Thus old source history supplies features, not labels from subsequent neutral trajectories or a human benchmark.

The fixed control pools these same features over the six earlier development sources and locks its 12 extended entries. The random control samples 12 entries with a frozen seed, whereas the full control supplies all 201 beyond the short common set. Every arm preserves the preceding compressed interface and runs the same upper codec after importing its chosen certificates. That codec can add further support definitions, so equal catalogue quotas do not imply equal final anchor counts or bytes.

\subsection*{F.4. Native source construction and permitted continuation}\label{f.4.-native-source-construction-and-permitted-continuation}
\addcontentsline{toc}{subsection}{F.4. Native source construction and permitted continuation}

The common source inherits the proof-closed 2,708-record library from Appendix E. Its elementary kernel checks types, contexts, distinct-variable conditions, and substitutions. The cumulative adapter retains the earlier premise-pair model on its supported pure implication/negation sublanguage; other native syntax receives the declared intrinsic fallback score. No experiment here validates learned transfer to arbitrary ZFC syntax.

During source construction, the fixed half-filter and deferred queue operate together. The queue holds at most 32,768 opportunities, and the constructor caps active derived premises at 128. The ordinary fragment selector reserves four positions. Rot-pruning deactivates a premise after eight idle blocks if that premise shows no earlier productive use and holds no protected lease. A block contains 16 ordinary opportunities. The upper codec preserves the accumulated interface and does not restrict it to the 128-place construction frontier.

The neutral continuer receives a physical proof-closure projection of the chosen interface. A certificate-only ancestor never supplies a premise or subsumer. It also cannot cause search to reject a newly derived statement solely because the archive already knows that statement. A fresh witness from allowed parents must pass native checking before the old statement acquires search rights; the export includes that witness under a separate identifier. Both the unique main archive and all new rediscovery witnesses pass Metamath.

Uniform continuation samples from allowed premises. Mixed continuation cycles through uniform, bridging, contracting, and frontier sampling in blocks of 16 opportunities. Both retain the existing generalisation sampler and perform the same native admissibility checks. They clear the state that the reuse mechanism and constructor pass on, retain newly admitted results, and apply no new human-conditioned choice. The fixed mixed policy depends on certificate identifiers, so its comparison does not establish representation-invariant utility.

\subsection*{F.5. Tested adaptive catalogue score}\label{f.5.-tested-adaptive-catalogue-score}
\addcontentsline{toc}{subsection}{F.5. Tested adaptive catalogue score}

The native confirmation fixes a common catalogue of 217 delivered schemata from programmes with at most three open MP nodes. Every arm receives the 16 delivered outputs of the short subgrammar with at most two nodes. The adaptive arm selects 12 additional outputs using only the new part of its source proof DAG. The quota matches the number of selected catalogue outputs, not total anchors, certificate support, or a universal compression threshold.

For each structural feature \(f\), the source profile counts correctly instantiated boundary-equality patterns and weights each occurrence by one plus the already observed direct uses of the fragment's root. Parent substitutions compose along the fragment before the profile compares boundary formulae; repeated node identifiers alone do not establish equality. If \(c_{j}(f)\) denotes this past profile, \(k(f)\) its MP-node count, and \(\mathcal{F}(\ell)\) the programme features associated with catalogue entry \(\ell\), the fixed score follows

\[\fitdisplay{s_{j}(\ell) = \frac{\sum_{f \in \mathcal{F}(\ell)}^{}k(f)log\left( 1 + c_{j}(f) \right)}{|\ell|_{AST}}.}\]

The score favours recurrent multistep patterns and charges for the symbols that describe them. A deterministic identifier order breaks ties. It estimates reuse from the source's past and neither trains on human targets nor predicts a causal reduction in future proving work. The fixed control applies the same ranking to pooled earlier development profiles and locks its 12 entries before the new sources run; random and whole-catalogue controls test two alternatives. The catalogue itself costs computation: every arm receives the full recorded cold catalogue charge.

\section*{Appendix G. Frozen native holdout and audit}\label{appendix-g.-frozen-native-holdout-and-audit}
\addcontentsline{toc}{section}{Appendix G. Frozen native holdout and audit}

\subsection*{G.1. Commitment before held-out outcomes}\label{g.1.-commitment-before-held-out-outcomes}
\addcontentsline{toc}{subsection}{G.1. Commitment before held-out outcomes}

The original v2 native package froze the corpus, source states, algorithms, catalogue, quotas, budgets, code, and decision rule before its search started. Its local seal supplies a content commitment, not an externally registered or authenticated timestamp. The single corpus commitment also prevents an ordinary launcher call from silently preparing another confirmation with the same target pool.

Input checks exclude the nine named-constant files before the seal and any confirmation trajectory. Technical checks use only the older reference and cannot contribute evidence of held-out superiority.

\subsection*{G.2. Corpus and exclusions}\label{g.2.-corpus-and-exclusions}
\addcontentsline{toc}{subsection}{G.2. Corpus and exclusions}

A strict parser reads theorem terms without executing TPTP or Prolog code and never extracts archive members into arbitrary paths. The parser admits universally quantified formulae over implication, negation, and wff variables, rejects unknown functions or free variables, and canonicalises variable names by first occurrence. Nine files with named truth or falsehood constants remain outside this declared language.

We compare the remaining formulae with every supplied statement in the earlier 919-target native pool after we expand the specified definitions. The earlier pool yields 484 distinct pure-propositional expansions; 407 rows fall outside that language, and further duplicate expansions explain the difference from 919. The procedure removes 402 POI entries and retains 963 distinct statements. No remaining entry exceeds the predeclared bound of 160 nodes per stored formula. It removes exact expanded overlap, not all semantically equivalent or substitution-related statements.

POI derives from the core of a July 2025 Metamath revision and converts essential hypotheses into implication antecedents. The native source uses its separately pinned explicit basis. Their common ancestry limits distributional independence, even though no retained POI target supplies a training label or an earlier recorded native score. The experiment tests previously unscored target formulae, not an unseen mathematical tradition or external-prover superiority.

\subsection*{G.3. Frozen arms and clocks}\label{g.3.-frozen-arms-and-clocks}
\addcontentsline{toc}{subsection}{G.3. Frozen arms and clocks}

The original attempt uses RNG states 2026102101--2026102116; the disclosed pre-scoring technical replacement uses 2026102201--2026102216. Each runs 1,500 cumulative opportunities over the common inherited library before any arm-specific rule choice. Every arm receives the original source interface and the 16 short catalogue entries. Core adds no extended entry; fixed adds the 12 entries that it selected on older profiles; random adds 12 seeded entries; adaptive adds the 12 entries with the highest source-profile scores; fixed-all adds every extended entry. The remaining upper coding and interface projection follow the same rule.

Within each source, we match RNG seeds across arms separately for the uniform and mixed continuers. Two local workers execute the fixed plan; all scientific processes fix the hash seed and numerical-library thread limits. Complete trajectories receive 90 growth CPU-seconds so that both endpoints, at 60 growth CPU-seconds and 90 cold operational CPU-seconds, remain observable. An atomic primitive can overrun a horizon, but a statement receives no credit at that horizon unless its recorded completion time fits.

The cold operational clock charges the CPU time that the pipeline spends before it resumes growth. The pipeline reads and extends the source, then synthesises and certifies the full catalogue. It prepares each method, imports rules, codes proofs, projects the interface, exports and verifies the initial package, and sets up the next run. Historical library construction and model training remain outside this account. Interpreter startup, filesystem waiting, the offline target observer, terminal bulk future export, and the independent audit also remain separate. The experiment consequently compares a declared incremental operational scope, not total research cost or a complete from-axioms proof job.

The formula bound constrains stored canonical conclusions at 160 nodes. The certificate-height bound equals 256, and the admission guard permits up to one million future events. We cannot claim complete-horizon superiority if a guard stops any cell early or any cell fails or remains missing. The frozen policy forbids outcome-guided extension, cancellation, or rerunning failed cells as ordinary confirmation.

\subsection*{G.4. Estimands and statistical decision}\label{g.4.-estimands-and-statistical-decision}
\addcontentsline{toc}{subsection}{G.4. Estimands and statistical decision}

Let \(t_{j,m,p}(u)\) denote the first growth CPU time at which a permitted statement covers target \(u\) for source \(j\), method \(m\), and continuer \(p\). Unreached targets carry infinite first-cover time. Define the two scores

\[\fitdisplay{q_{j,m}^{future} = \frac{1}{2}\sum_{p = 1}^{2}{\sum_{u \in R_{j}}^{}\mathbf{1}}\{ t_{j,m,p}(u) \leq 60\},}\]

\[\fitdisplay{q_{j,m}^{cold} = \frac{1}{2}\sum_{p = 1}^{2}{\sum_{u \in R_{j}}^{}\mathbf{1}}\{ c_{j,m,p} + t_{j,m,p}(u) \leq 90\},}\]

where \(c_{j,m,p}\) records the declared pre-growth operational cost. The common residual \(R_{j}\) removes targets that any arm initially covers. Exact matches, coverage fractions, first-cover curve areas, costs, and interface sizes supply secondary descriptions.

For each control \(a\) and endpoint \(z\), paired differences follow \(d_{j}^{z} = q_{j,adaptive}^{z} - q_{j,a}^{z}\). The exact one-sided sign-flip tail equals

\[\fitdisplay{p_{a}^{z} = 2^{- 16}\sum_{\epsilon \in \{ - 1,1\}^{16}}^{}\mathbf{1}\left\{ \sum_{j}^{}\epsilon_{j}d_{j}^{z} \geq \sum_{j}^{}d_{j}^{z} \right\}.}\]

This test assumes exchangeability of paired signs under its null; the source states do not sample independent axiom systems. The joint comparison uses \(p_{a} = max(p_{a}^{future},p_{a}^{cold})\), then Holm's step-down correction across fixed, random, and fixed-all. Confirmation requires every adjusted p-value below 0.05 and every paired mean difference at least one target at both endpoints. A nonempty common residual, all planned observations, varying adaptive selections, and a passing independent audit remain additional conditions.

Zero or nearly identical residual coverage cannot rank semantic quality. If a method fails the rule, this test does not confirm adaptive superiority; it establishes neither equivalence nor the impossibility of useful adaptation. We retain the locked outcome and do not choose new weights on these targets.

\subsection*{G.5. Audit procedure and the original pre-scoring interruption}\label{g.5.-audit-procedure-and-the-original-pre-scoring-interruption}
\addcontentsline{toc}{subsection}{G.5. Audit procedure and the original pre-scoring interruption}

Blind workers read only committed code, construction inputs, and protocol metadata; they do not read target bytes. A receipt hashes every completed source block. The evaluator verifies all 16 receipts before it creates an exclusive opening record, loads targets, and observes previously committed trajectories. No interim target score can control the search. This separation enforces the software workflow; it does not impose an operating-system barrier against deliberate manual access.

The evaluator records exact and one-way first hits, covering statements, and explicit substitution witnesses. The evaluator supplies an instance certificate for every credited target so the independent Metamath verifier can check it. The auditor separately reconstructs permitted premises, new hidden-support witnesses, initial interface projection, and upper coding; it recomputes target-oriented earliest hits independently of the producer's aggregate counts. It then rebuilds each residual set and both endpoints, recomputes the paired tests and correction, and checks the scientific conclusion against all eligibility gates.

\subsection*{G.6. Disclosed technical amendment before scoring}\label{g.6.-disclosed-technical-amendment-before-scoring}
\addcontentsline{toc}{subsection}{G.6. Disclosed technical amendment before scoring}

The repaired importer computes a descendant's axiom origins from the destination parents that its certificate actually cites. Canonical duplicates retain their existing first proof. The patch changes neither the checking kernel nor the logical axiom base, and every imported step must still satisfy all proof, substitution, and distinct-variable checks. A new regression test supplies identical axiom schemata under different source labels and verifies the merged result through official Metamath.

The replacement package retains the same unopened 963-target pool, hypothesis, model, inherited library, rule catalogue, quotas, clocks, controls, and success rule, but uses fresh source states 2026102201--2026102216. It imports no earlier blind trajectory or score. The four completed blocks from the failed attempt remain outside its sample. The replacement subsequently completes all 16 blocks and 160 continuations; the pipeline then evaluates the results once and runs an independent audit. It follows a technical failure before the first target score; it does not supply an independent replication or a new mathematical corpus. Section G.7 records its actual scientific outcome.

\subsection*{G.7. Completed comparison and independent audit}\label{g.7.-completed-comparison-and-independent-audit}
\addcontentsline{toc}{subsection}{G.7. Completed comparison and independent audit}

The corrected run completes all 16 blind blocks and 160 continuations before the evaluator opens the 963 targets. Both registered horizons finish in every cell, and the common residual remains informative. The selector supplies three distinct extended-rule sets. Its scientific outcome follows the frozen rule: no confirmed adaptive superiority (Table S3).

\begingroup
\small
\setlength{\tabcolsep}{3pt}
\renewcommand{\arraystretch}{1.13}

\begin{longtable}[]{@{}
  >{\raggedright\arraybackslash}p{(\linewidth - 8\tabcolsep) * \real{0.2558}}
  >{\raggedleft\arraybackslash}p{(\linewidth - 8\tabcolsep) * \real{0.1701}}
  >{\raggedleft\arraybackslash}p{(\linewidth - 8\tabcolsep) * \real{0.1865}}
  >{\raggedleft\arraybackslash}p{(\linewidth - 8\tabcolsep) * \real{0.2259}}
  >{\raggedleft\arraybackslash}p{(\linewidth - 8\tabcolsep) * \real{0.1617}}@{}}
\caption{Full locked native comparison across 16 source blocks and both registered horizons}\label{tab:34}\tabularnewline
\toprule\noalign{}
\begin{minipage}[b]{\linewidth}\raggedright
\textbf{Library}
\end{minipage} & \begin{minipage}[b]{\linewidth}\raggedleft
\textbf{Initial coverage}
\end{minipage} & \begin{minipage}[b]{\linewidth}\raggedleft
\textbf{New at 60 growth s}
\end{minipage} & \begin{minipage}[b]{\linewidth}\raggedleft
\textbf{New at 90 operational s}
\end{minipage} & \begin{minipage}[b]{\linewidth}\raggedleft
\textbf{Derived anchors}
\end{minipage} \\
\midrule\noalign{}
\endfirsthead
\toprule\noalign{}
\begin{minipage}[b]{\linewidth}\raggedright
\textbf{Library}
\end{minipage} & \begin{minipage}[b]{\linewidth}\raggedleft
\textbf{Initial coverage}
\end{minipage} & \begin{minipage}[b]{\linewidth}\raggedleft
\textbf{New at 60 growth s}
\end{minipage} & \begin{minipage}[b]{\linewidth}\raggedleft
\textbf{New at 90 operational s}
\end{minipage} & \begin{minipage}[b]{\linewidth}\raggedleft
\textbf{Derived anchors}
\end{minipage} \\
\midrule\noalign{}
\endhead
\bottomrule\noalign{}
\endlastfoot
Common core and short rules & 25.00 & 1.312 & 1.344 & 633.6 \\
Fixed extended repertoire & 29.00 & 1.594 & 1.625 & 675.0 \\
Random extended repertoire & 26.38 & 1.500 & 1.562 & 680.4 \\
Adaptive test variant & 27.88 & 1.656 & 1.688 & 673.4 \\
Whole fixed catalogue & 44.00 & 1.562 & 1.594 & 1282.3 \\
\end{longtable}

\endgroup

Table S4 reports the paired tests for the registered adaptive comparisons.

\begingroup
\small
\setlength{\tabcolsep}{3pt}
\renewcommand{\arraystretch}{1.13}

\begin{longtable}[]{@{}
  >{\raggedright\arraybackslash}p{(\linewidth - 6\tabcolsep) * \real{0.3000}}
  >{\raggedleft\arraybackslash}p{(\linewidth - 6\tabcolsep) * \real{0.2400}}
  >{\raggedleft\arraybackslash}p{(\linewidth - 6\tabcolsep) * \real{0.2400}}
  >{\raggedleft\arraybackslash}p{(\linewidth - 6\tabcolsep) * \real{0.2200}}@{}}
\caption{Full paired adaptive-minus-control tests under the registered joint-endpoint rule}\label{tab:35}\tabularnewline
\toprule\noalign{}
\begin{minipage}[b]{\linewidth}\raggedright
\textbf{Adaptive minus control}
\end{minipage} & \begin{minipage}[b]{\linewidth}\raggedleft
\textbf{Difference at 60 growth s}
\end{minipage} & \begin{minipage}[b]{\linewidth}\raggedleft
\textbf{Difference at 90 operational s}
\end{minipage} & \begin{minipage}[b]{\linewidth}\raggedleft
\textbf{Holm adjusted p}
\end{minipage} \\
\midrule\noalign{}
\endfirsthead
\toprule\noalign{}
\begin{minipage}[b]{\linewidth}\raggedright
\textbf{Adaptive minus control}
\end{minipage} & \begin{minipage}[b]{\linewidth}\raggedleft
\textbf{Difference at 60 growth s}
\end{minipage} & \begin{minipage}[b]{\linewidth}\raggedleft
\textbf{Difference at 90 operational s}
\end{minipage} & \begin{minipage}[b]{\linewidth}\raggedleft
\textbf{Holm adjusted p}
\end{minipage} \\
\midrule\noalign{}
\endhead
\bottomrule\noalign{}
\endlastfoot
fixed & +0.06250 & +0.06250 & 1.000 \\
random & +0.15625 & +0.12500 & 1.000 \\
fixed all & +0.09375 & +0.09375 & 1.000 \\
\end{longtable}

\endgroup

At 60 growth seconds, the adaptive variant records 3 wins, 11 ties, and 2 losses against fixed; 7 wins, 2 ties, and 7 losses against random; and 6 wins, 4 ties, and 6 losses against fixed-all. The corresponding unadjusted one-sided p-values equal 0.500000, 0.295654, and 0.380371. At the cold endpoint they equal 0.500000, 0.347534, and 0.377686. Taking the larger p across clocks and applying Holm yields 1.000 for every comparison. Positive descriptive means do not meet the required mean margin or multiplicity rule.

All 16 source blocks and 160 cells pass the independent checker. It reconstructs 4,174,478 admission events and 2,846 rediscovery witnesses, checks all 963 target positions, and recomputes permissions, projected interfaces, earliest cover times, residuals, tests, and the conclusion. The event count includes repeated trajectory events, not that many independent statements (Fig. S1).

\begin{figure}[!htbp]
\centering
\includegraphics[width=\linewidth,height=.67\textheight,keepaspectratio]{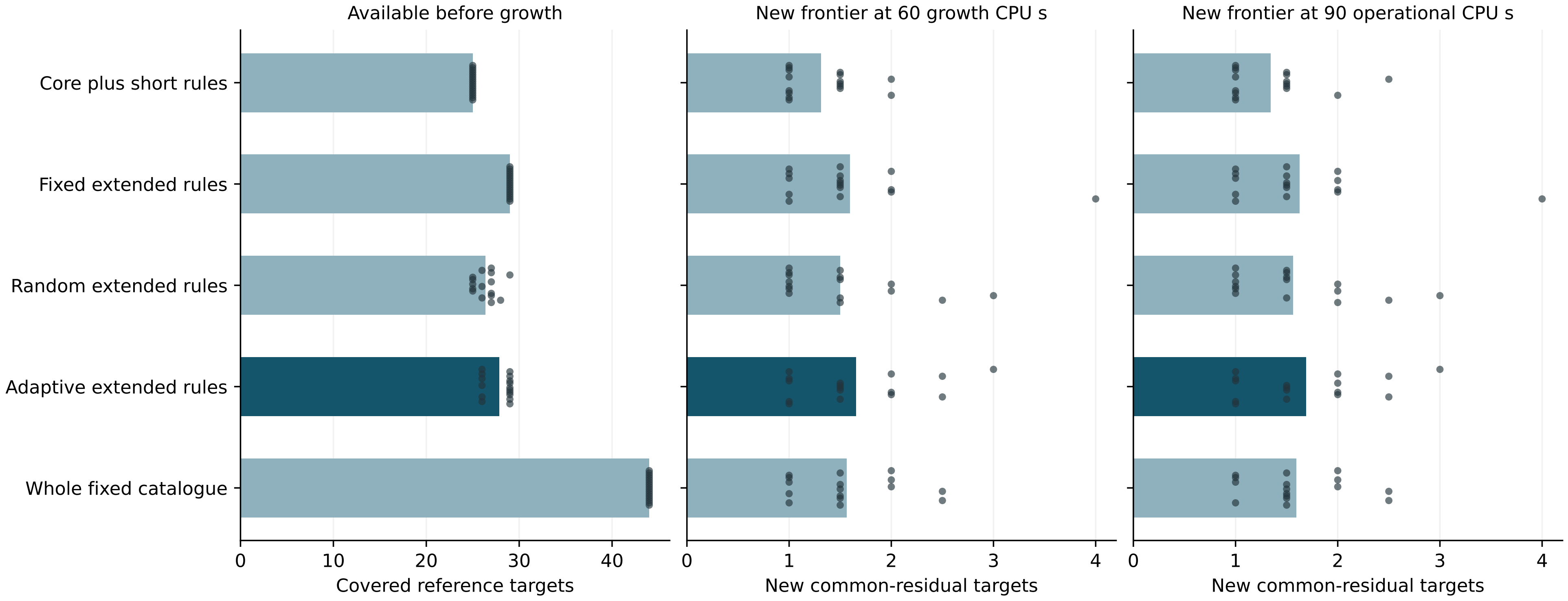}
\caption{Locked native comparison. Bars represent means over 16 source blocks; dots show source means after averaging the two neutral continuers. Initial coverage counts the whole reference, whereas the future panels count the common initial residual. The figure supplies descriptive variation, not confidence intervals}\label{fig:12}
\end{figure}

The frozen package contains the full registered report for every arm and paired outcome. We neither retune on these targets nor infer equivalence from the failed superiority test. The fixed repertoire supplies the working configuration, not a post-hoc claim that it beats every alternative.

\section*{Appendix H. Cumulative skeleton networks}\label{appendix-h.-cumulative-skeleton-networks}
\addcontentsline{toc}{section}{Appendix H. Cumulative skeleton networks}

\subsection*{H.1. Projections and representative choice}\label{h.1.-projections-and-representative-choice}
\addcontentsline{toc}{subsection}{H.1. Projections and representative choice}

The new descriptive study reads only audited initial libraries from the corrected holdout. Each projection includes every available anchor, all 29 basis roots, and every necessary certificate ancestor. The projection joins each parent to the conclusion and collapses duplicate arcs. Hidden ancestors retain proof roles but no automatic search rights. The first source, 2026102201, supplies both representatives. The main text uses its fixed quota-matched control because adaptive ranking failed the registered rule; the image below retains the adaptive representative that we chose before scoring. This editorial fallback does not relabel adaptive as an effective method or select the highest-performing control.

Panels a, c, and d retain their old graph hashes and layouts. Panels c and d also retain their shared coordinates, colours, and vertex areas. Panel b uses the same fixed layout seed, three-start community search, and vertex-area rule. All vertices and unique undirected edges enter every drawing. Because each panel uses its own scale, distances across panels do not measure comparable semantic proximity (Fig. S2).

\begin{figure}[!htbp]
\centering
\includegraphics[width=\linewidth,height=.67\textheight,keepaspectratio]{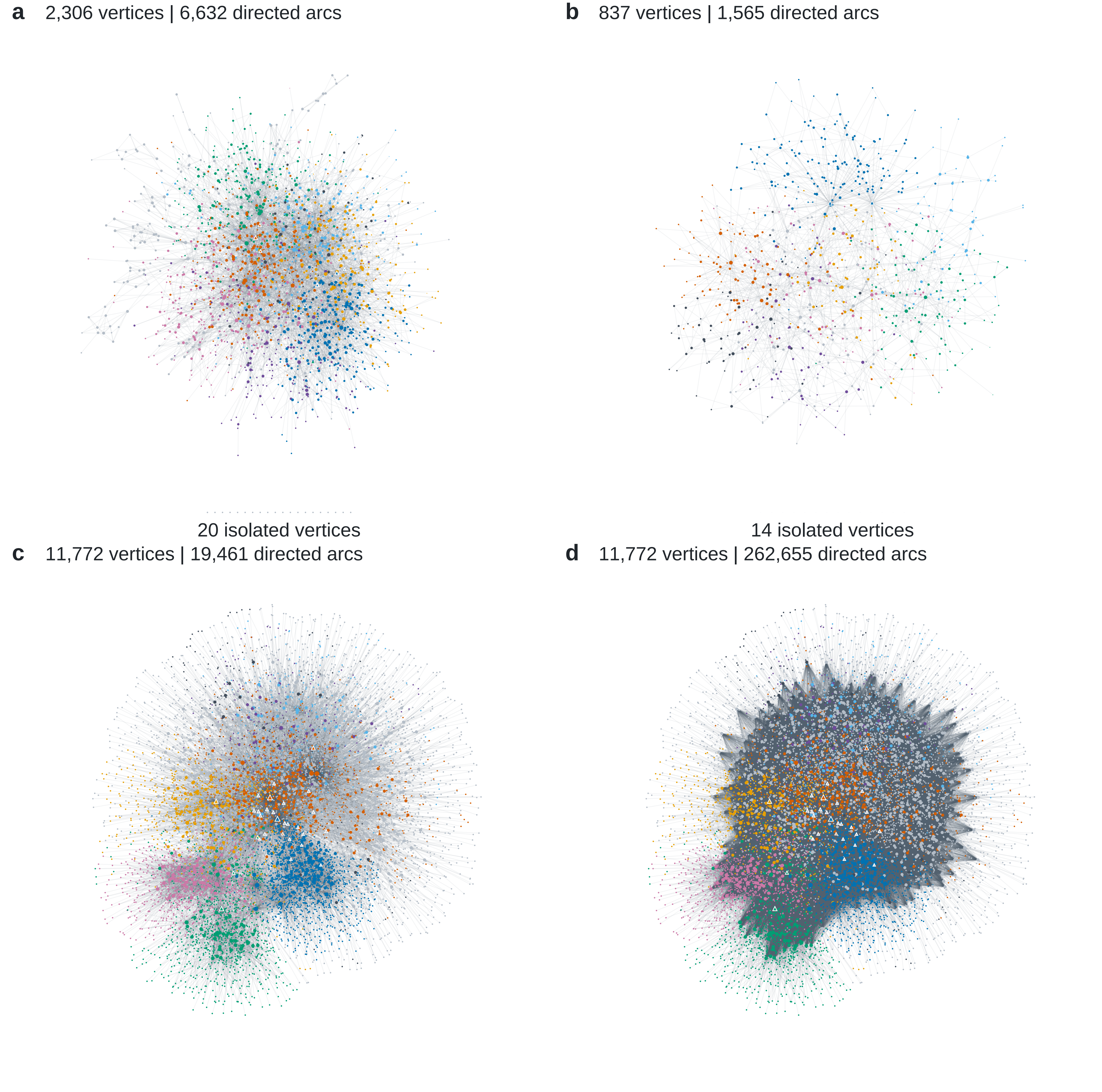}
\caption{Preselected adaptive test variant on the same basis. Panel b uses adaptive initial source 2026102201, not the best continuation or target-coverage result. The remaining panels retain the frozen Metamath and height-three graphs and layouts. This graph illustrates a tested variant without an established adaptive advantage}\label{fig:14}
\end{figure}

\subsection*{H.2. Profiles across all source states}\label{h.2.-profiles-across-all-source-states}
\addcontentsline{toc}{subsection}{H.2. Profiles across all source states}

Every source contributes one initial library for each of the two reported methods, not two independent theories. Anchors exclude basis roots; certificate vertices include roots and hidden support. C denotes mean local clustering, L the exact mean path on the largest undirected component, Q the best modularity across three Louvain starts, and Q0 the mean over ten simple degree-preserving null graphs. For every new observed and null graph, we compute paths from every BFS source.

\subsubsection*{Fixed repertoire}\label{fixed-repertoire}
\addcontentsline{toc}{subsubsection}{Fixed repertoire}

Table S5 reports each fixed-repertoire initial skeleton; source suffixes follow the frozen order.

\begingroup
\small
\setlength{\tabcolsep}{3pt}
\renewcommand{\arraystretch}{1.13}

\begin{longtable}[]{@{}
  >{\raggedright\arraybackslash}p{(\linewidth - 14\tabcolsep) * \real{0.1238}}
  >{\raggedleft\arraybackslash}p{(\linewidth - 14\tabcolsep) * \real{0.1172}}
  >{\raggedleft\arraybackslash}p{(\linewidth - 14\tabcolsep) * \real{0.1732}}
  >{\raggedleft\arraybackslash}p{(\linewidth - 14\tabcolsep) * \real{0.1172}}
  >{\raggedleft\arraybackslash}p{(\linewidth - 14\tabcolsep) * \real{0.1172}}
  >{\raggedleft\arraybackslash}p{(\linewidth - 14\tabcolsep) * \real{0.1172}}
  >{\raggedleft\arraybackslash}p{(\linewidth - 14\tabcolsep) * \real{0.1172}}
  >{\raggedleft\arraybackslash}p{(\linewidth - 14\tabcolsep) * \real{0.1172}}@{}}
\caption{Network profiles of all 16 fixed-repertoire initial skeletons}\label{tab:39}\tabularnewline
\toprule\noalign{}
\begin{minipage}[b]{\linewidth}\raggedright
\textbf{Source suffix}
\end{minipage} & \begin{minipage}[b]{\linewidth}\raggedleft
\textbf{Anchors}
\end{minipage} & \begin{minipage}[b]{\linewidth}\raggedleft
\textbf{Certificate vertices}
\end{minipage} & \begin{minipage}[b]{\linewidth}\raggedleft
\textbf{Arcs}
\end{minipage} & \begin{minipage}[b]{\linewidth}\raggedleft
\textbf{C}
\end{minipage} & \begin{minipage}[b]{\linewidth}\raggedleft
\textbf{L}
\end{minipage} & \begin{minipage}[b]{\linewidth}\raggedleft
\textbf{Q}
\end{minipage} & \begin{minipage}[b]{\linewidth}\raggedleft
\textbf{Q0}
\end{minipage} \\
\midrule\noalign{}
\endfirsthead
\toprule\noalign{}
\begin{minipage}[b]{\linewidth}\raggedright
\textbf{Source suffix}
\end{minipage} & \begin{minipage}[b]{\linewidth}\raggedleft
\textbf{Anchors}
\end{minipage} & \begin{minipage}[b]{\linewidth}\raggedleft
\textbf{Certificate vertices}
\end{minipage} & \begin{minipage}[b]{\linewidth}\raggedleft
\textbf{Arcs}
\end{minipage} & \begin{minipage}[b]{\linewidth}\raggedleft
\textbf{C}
\end{minipage} & \begin{minipage}[b]{\linewidth}\raggedleft
\textbf{L}
\end{minipage} & \begin{minipage}[b]{\linewidth}\raggedleft
\textbf{Q}
\end{minipage} & \begin{minipage}[b]{\linewidth}\raggedleft
\textbf{Q0}
\end{minipage} \\
\midrule\noalign{}
\endhead
\bottomrule\noalign{}
\endlastfoot
01 & 675 & 841 & 1573 & 0.0202 & 4.177 & 0.622 & 0.539 \\
02 & 679 & 852 & 1594 & 0.0199 & 4.179 & 0.622 & 0.542 \\
03 & 665 & 836 & 1564 & 0.0189 & 4.175 & 0.623 & 0.541 \\
04 & 669 & 840 & 1572 & 0.0200 & 4.175 & 0.624 & 0.541 \\
05 & 679 & 847 & 1585 & 0.0200 & 4.167 & 0.623 & 0.541 \\
06 & 669 & 840 & 1571 & 0.0199 & 4.171 & 0.622 & 0.543 \\
07 & 682 & 848 & 1586 & 0.0186 & 4.169 & 0.628 & 0.541 \\
08 & 672 & 844 & 1578 & 0.0207 & 4.179 & 0.623 & 0.541 \\
09 & 674 & 845 & 1583 & 0.0215 & 4.181 & 0.624 & 0.541 \\
10 & 677 & 849 & 1587 & 0.0207 & 4.177 & 0.629 & 0.541 \\
11 & 669 & 840 & 1571 & 0.0188 & 4.183 & 0.625 & 0.540 \\
12 & 674 & 848 & 1589 & 0.0203 & 4.180 & 0.628 & 0.541 \\
13 & 681 & 848 & 1586 & 0.0186 & 4.189 & 0.625 & 0.544 \\
14 & 681 & 849 & 1588 & 0.0207 & 4.165 & 0.624 & 0.540 \\
15 & 672 & 844 & 1579 & 0.0204 & 4.175 & 0.629 & 0.539 \\
16 & 682 & 851 & 1594 & 0.0201 & 4.171 & 0.624 & 0.542 \\
\end{longtable}

\endgroup

Mean anchors: 675.0000; mean certificate vertices: 845.1250; mean modularity: 0.6246.

\subsubsection*{Adaptive test variant}\label{adaptive-test-variant}
\addcontentsline{toc}{subsubsection}{Adaptive test variant}

Table S6 reports the adaptive-test initial skeletons in the same source order.

\begingroup
\small
\setlength{\tabcolsep}{3pt}
\renewcommand{\arraystretch}{1.13}

\begin{longtable}[]{@{}
  >{\raggedright\arraybackslash}p{(\linewidth - 14\tabcolsep) * \real{0.1238}}
  >{\raggedleft\arraybackslash}p{(\linewidth - 14\tabcolsep) * \real{0.1172}}
  >{\raggedleft\arraybackslash}p{(\linewidth - 14\tabcolsep) * \real{0.1732}}
  >{\raggedleft\arraybackslash}p{(\linewidth - 14\tabcolsep) * \real{0.1172}}
  >{\raggedleft\arraybackslash}p{(\linewidth - 14\tabcolsep) * \real{0.1172}}
  >{\raggedleft\arraybackslash}p{(\linewidth - 14\tabcolsep) * \real{0.1172}}
  >{\raggedleft\arraybackslash}p{(\linewidth - 14\tabcolsep) * \real{0.1172}}
  >{\raggedleft\arraybackslash}p{(\linewidth - 14\tabcolsep) * \real{0.1172}}@{}}
\caption{Network profiles of all 16 adaptive-test initial skeletons}\label{tab:40}\tabularnewline
\toprule\noalign{}
\begin{minipage}[b]{\linewidth}\raggedright
\textbf{Source suffix}
\end{minipage} & \begin{minipage}[b]{\linewidth}\raggedleft
\textbf{Anchors}
\end{minipage} & \begin{minipage}[b]{\linewidth}\raggedleft
\textbf{Certificate vertices}
\end{minipage} & \begin{minipage}[b]{\linewidth}\raggedleft
\textbf{Arcs}
\end{minipage} & \begin{minipage}[b]{\linewidth}\raggedleft
\textbf{C}
\end{minipage} & \begin{minipage}[b]{\linewidth}\raggedleft
\textbf{L}
\end{minipage} & \begin{minipage}[b]{\linewidth}\raggedleft
\textbf{Q}
\end{minipage} & \begin{minipage}[b]{\linewidth}\raggedleft
\textbf{Q0}
\end{minipage} \\
\midrule\noalign{}
\endfirsthead
\toprule\noalign{}
\begin{minipage}[b]{\linewidth}\raggedright
\textbf{Source suffix}
\end{minipage} & \begin{minipage}[b]{\linewidth}\raggedleft
\textbf{Anchors}
\end{minipage} & \begin{minipage}[b]{\linewidth}\raggedleft
\textbf{Certificate vertices}
\end{minipage} & \begin{minipage}[b]{\linewidth}\raggedleft
\textbf{Arcs}
\end{minipage} & \begin{minipage}[b]{\linewidth}\raggedleft
\textbf{C}
\end{minipage} & \begin{minipage}[b]{\linewidth}\raggedleft
\textbf{L}
\end{minipage} & \begin{minipage}[b]{\linewidth}\raggedleft
\textbf{Q}
\end{minipage} & \begin{minipage}[b]{\linewidth}\raggedleft
\textbf{Q0}
\end{minipage} \\
\midrule\noalign{}
\endhead
\bottomrule\noalign{}
\endlastfoot
01 & 668 & 837 & 1565 & 0.0197 & 4.170 & 0.631 & 0.541 \\
02 & 679 & 852 & 1594 & 0.0199 & 4.179 & 0.622 & 0.542 \\
03 & 662 & 832 & 1556 & 0.0184 & 4.169 & 0.626 & 0.542 \\
04 & 673 & 843 & 1578 & 0.0185 & 4.160 & 0.624 & 0.541 \\
05 & 672 & 843 & 1577 & 0.0195 & 4.161 & 0.626 & 0.542 \\
06 & 669 & 840 & 1571 & 0.0199 & 4.171 & 0.622 & 0.543 \\
07 & 682 & 848 & 1586 & 0.0186 & 4.169 & 0.628 & 0.541 \\
08 & 669 & 840 & 1570 & 0.0202 & 4.173 & 0.619 & 0.541 \\
09 & 674 & 845 & 1583 & 0.0215 & 4.181 & 0.624 & 0.541 \\
10 & 677 & 849 & 1587 & 0.0207 & 4.177 & 0.629 & 0.541 \\
11 & 669 & 840 & 1571 & 0.0188 & 4.183 & 0.625 & 0.540 \\
12 & 674 & 848 & 1589 & 0.0203 & 4.180 & 0.628 & 0.541 \\
13 & 675 & 846 & 1583 & 0.0181 & 4.182 & 0.619 & 0.542 \\
14 & 681 & 849 & 1588 & 0.0207 & 4.165 & 0.624 & 0.540 \\
15 & 669 & 840 & 1571 & 0.0199 & 4.169 & 0.617 & 0.540 \\
16 & 682 & 851 & 1594 & 0.0201 & 4.171 & 0.624 & 0.542 \\
\end{longtable}

\endgroup

Mean anchors: 673.4375; mean certificate vertices: 843.9375; mean modularity: 0.6243.

\subsection*{H.3. Tail diagnostics and reproducibility}\label{h.3.-tail-diagnostics-and-reproducibility}
\addcontentsline{toc}{subsection}{H.3. Tail diagnostics and reproducibility}

For each outgoing distinct-use profile, the diagnostic uses 20,000 bootstrap draws for the exact discrete pure power-law family and refits the threshold and exponent in every draw. The earlier in-degree and undirected-degree alternatives remain secondary descriptions. Conditional tail diagnostics do not assume independent graph replicates or establish a limiting law. Small-world ratios compare clustering and largest-component paths with ten degree-preserving controls, not a universal classifier.

Across all 16 fixed sources, pure-family bootstrap p-values range from 0.24190 to 0.62575; across the 16 adaptive variants, they range from 0.17950 to 0.62575. None rejects that family at five percent. The new certificate projections retain 13--15 isolated unused basis roots, which still enter the vertex count and zero-degree mass; both illustrated representatives contain 14. A compatible finite tail neither confirms adaptive future-use superiority nor proves that the generator intrinsically converges to a scale-free network. The machine-readable profiles retain thresholds, exponents, binomial Monte Carlo intervals, zero-degree counts, and same-tail alternative likelihoods for every source.

\clearpage
\section*{Appendix I. The propositional skeleton and its external proofs}
\addcontentsline{toc}{section}{Appendix I. The propositional skeleton and its external proofs}

\subsection*{I.1 The ten anchors and their common certificate}
\addcontentsline{toc}{subsection}{I.1 The ten anchors and their common certificate}

Rows give the selection order at seed 101 and the node identifiers. Implication associates to the right. Variables are local to each schema.

\begin{center}\small
\begin{tabular}{@{}rrl@{}}\toprule
Rank & Node & Anchor schema \\ \midrule
1 & 6 & \((A\to B\to C\to D)\to A\to (B\to C)\to B\to D\) \\
2 & 107 & \(A\to B\to C\to A\) \\
3 & 203 & \(A\to B\to B\) \\
4 & 8 & \(((A\to B\to C)\to ((A\to B)\to A\to C)\to D)\to (A\to B\to C)\to D\) \\
5 & 167 & \(A\to A\) \\
6 & 32 & \(A\to B\to C\to D\to C\) \\
7 & 295 & \((A\to \neg B\to \neg C)\to A\to C\to B\) \\
8 & 65 & \((A\to (B\to A)\to C)\to A\to C\) \\
9 & 40 & \((A\to A\to B)\to A\to B\) \\
10 & 45 & \(A\to B\to (\neg C\to \neg D)\to D\to C\) \\
\bottomrule\end{tabular}
\end{center}

Start with nodes \(0,1,2\), respectively,
\[
\begin{aligned}
0&:\ A\to(B\to A),\\
1&:\ (A\to(B\to C))\to((A\to B)\to(A\to C)),\\
2&:\ (\neg A\to\neg B)\to(B\to A).
\end{aligned}
\]
Read down each column pair, starting on the left. Rename the two premises' variables apart and unify the major antecedent with the minor premise. The most general unifier applied to the major consequent yields the node's schema. The table omits archive steps outside the certificate.

\begin{center}\small
\begin{tabular}{@{}rl@{\qquad}rl@{\qquad}rl@{}}\toprule
Node & (Major, minor) & Node & (Major, minor) & Node & (Major, minor)\\ \midrule
3 & (0, 1) & 29 & (13, 3) & 72 & (31, 30) \\
4 & (0, 0) & 30 & (27, 27) & 106 & (72, 4) \\
5 & (1, 1) & 31 & (6, 27) & 107 & (65, 4) \\
6 & (1, 3) & 32 & (0, 4) & 125 & (1, 106) \\
8 & (5, 3) & 37 & (29, 32) & 167 & (106, 65) \\
10 & (8, 3) & 40 & (5, 27) & 178 & (52, 107) \\
13 & (10, 6) & 45 & (0, 28) & 188 & (125, 178) \\
27 & (1, 0) & 52 & (37, 3) & 203 & (52, 188) \\
28 & (0, 2) & 65 & (5, 4) & 295 & (1, 28) \\
\bottomrule\end{tabular}
\end{center}

The proof of \texttt{finch} uses nodes 6, 107, and 167. It retains nine of 16 steps, including the final inference. SGCD's full bank retains ten.

\clearpage
\subsection*{I.2 External panels and seed-level results}
\addcontentsline{toc}{subsection}{I.2 External panels and seed-level results}

We skip six argument-free aliases and translate the remaining 43 explicit definitions. Grouping their declared types up to variable renaming leaves 41 distinct targets. Before scoring external proofs we also specified a diagnostic subset by exact proof-graph hash. Thirty proofs differ from the earlier scored proofs and the proof structures emitted in the current and earlier 90-second comparison runs. Four of those 30 also have a canonical statement absent from our earlier statement corpora and growth archives. The three Aviary columns therefore overlap.

The seven published SGCD logs yield 2,633 proof records. We exclude 23 records outside the eligible target list, 127 previously scored proofs, and 337 proofs present in the growth archives. Across the remainder, the filter discards 435 repeated proof graphs and 1,236 further proofs of already represented statements. It chooses one proof per canonical target by the lowest proof hash, leaving 475 proofs of familiar statements. Distinct whole-proof hashes can share intermediate steps.

\begin{center}\small
\begin{tabular}{@{}lrrrrrr@{}}\toprule
& & & \multicolumn{3}{c}{Aviary} & Published\\
Run & Anchors & Cost & All 41 & New proofs (30) & New types (4) & SGCD (475)\\ \midrule
101 & 10 & 27 & 104 & 90 & 9 & 1,180 \\
202 & 13 & 22 & 137 & 121 & 13 & 1,305 \\
303 & 14 & 38 & 150 & 132 & 15 & 856 \\
404 & 23 & 30 & 148 & 130 & 15 & 1,308 \\
505 & 18 & 39 & 162 & 141 & 16 & 1,636 \\
SGCD & 22 & 28 & 146 & 128 & 15 & 1,308 \\
\bottomrule\end{tabular}
\end{center}

Cost counts the distinct inferences in a bank's shared certificate. The last four columns report gross inference steps saved, summed over proofs. Subtract cost once from any one column to obtain the net saving on that panel. The three deterministic SGCD runs return identical scores under the common growth ceiling of 90 CPU-seconds.

The ancillary file \texttt{skeleton-data.json} records both budget conditions. It includes the ordered anchors and shared certificate for each bank, the external proof graphs, and the anchor identifiers that each compressed proof uses. A reader can reconstruct the source and shortened step counts from these data.